\documentclass[%
superscriptaddress,
showpacs,preprintnumbers,
 amsmath,amssymb,
 aps,
 pre,
 longbibliography,
 lengthcheck,%
]{revtex4-1}

\usepackage{graphicx}
\usepackage{dcolumn}
\usepackage{bm}
\usepackage{hyperref}
\graphicspath{{Figures/}}

\begin{document}

\title{Effective diffusion coefficient in tilted disordered potentials: Optimal relative diffusivity at a finite temperature}

\author{R. Salgado-Garc\'{\i}a} 
\email{raulsg@uaem.mx}
\affiliation{Facultad de Ciencias, Universidad Aut\'onoma del Estado de Morelos. Avenida Universidad 1001, Colonia Chamilpa, 62209, Cuernavaca Morelos, Mexico.
} 

\date{\today}

\begin{abstract}

In this work we study the transport properties of non-interacting overdamped particles, moving on tilted disordered potentials, subjected to Gaussian white noise. We give exact formulas for the drift and diffusion coefficients for the case of random potentials resulting from the interaction of a particle with a ``random polymer''. In our model the random polymer is made up, by means of some stochastic process, of monomers that can be taken from a finite or countable infinite set of possible monomer types. For the case of uncorrelated random polymers we found that the diffusion coefficient exhibits a non-monotonous behavior as a function of the noise intensity. Particularly interesting is the fact that the relative diffusivity becomes optimal at a finite temperature, a behavior which is reminiscent of stochastic resonance. We explain this effect  as an interplay between the deterministic and noisy dynamics of the system. We also show that this behavior of the diffusion coefficient at a finite temperature is more pronounced for the case of weakly disordered potentials. We test our findings by means of numerical simulations of the corresponding Langevin dynamics of an ensemble of noninteracting overdamped particles diffusing on uncorrelated random potentials. 

\end{abstract}

\pacs{05.40.-a,05.60.-k,05.10.Gg}

\maketitle

\section{Introduction}
\label{sec:Intro}

It has been recognized that thermal diffusion of particles in one-dimensional (1D) potentials plays an important role in describing several physical systems, both at the mesoscale and at the nanoscale~\cite{risken1984fokker,hanggi2009artificial,brangwynne2009intracellular,bressloff2013stochastic}. Particularly, much effort has been done to understand several phenomena occurring in tilted periodic potentials, such as the giant enhancement of diffusion~\cite{reimann2001giant,reimann2002diffusion,evstigneev2008diffusion} or the enhancement of transport coherence~\cite{lindner2001optimal,lindner2002noise,dan2002giant,heinsalu2004diffusion}. These characteristics have become important because of its potential applications for technological purposes, such as particle separation~\cite{reimann2001giant,reimann2002diffusion,evstigneev2008diffusion}, DNA electrophoresis~\cite{viovy2000electrophoresis} or novel sequencing techniques~\cite{branton2008potential,ashkenasy2005recognizing}. Moreover, understanding the physics of thermal diffusion  would shed some light about several biological process involving 1D diffusion such as intracellular protein transport~\cite{bressloff2013stochastic,brangwynne2009intracellular,bressloff2013stochastic},  diffusion of proteins along DNA~\cite{slutsky2004diffusion,barsky2011proteins,shimamoto1999one,gorman2008visualizing,wang2006single}, or DNA translocation through a nanopore~\cite{slutsky2004diffusion,branton2008potential,ashkenasy2005recognizing}.  

The thermal diffusion of particles in 1D disordered potentials has also been the subject of intense research. Its importance lies on the fact that this class of systems has a diversity of behaviors which are not present in absence of disorder~\cite{bouchaud1990anomalous,haus1987diffusion,havlin1987diffusion}. For example, one of the earliest attempts to understand the thermal diffusion on one-dimensional disordered lattices is due to Sinai~\cite{sinai1983limiting}.  The so called Sinai's model has attracted much attention because it can be treated analytically~\cite{sinai1983limiting,solomon1975random,oshanin1993steady,burlatsky1992non} and represents one of the most simple models exhibiting several characteristics found in more complex systems such as anomalous diffusion~\cite{derrida1982classical,golosov1984localization,monthus2006random,monthus2003localization}. Another model that has also been widely explored corresponds to a system of overdamped particles moving on potentials with ``Gaussian disorder''~\cite{bouchaud1990classical,bouchaud1990anomalous,romero1998brownian,lopatin2001instanton,reimann2008weak,khoury2011weak}. It has been shown that this model exhibits normal and anomalous diffusion as well as normal and anomalous drift~\cite{bouchaud1990classical,romero1998brownian,lopatin2001instanton,pryadko2005driven}. In particular, this system has been used to understand some transport properties of proteins moving on DNA or the translocation of DNA thorough a nanopore~\cite{slutsky2004diffusion}. A third class of transport on random media  models are those with purely deterministic (overdamped or underdamped) dynamics~\cite{kunz2003mechanical,denisov2010biased,denisov2007analytically,denisov2007arrival,denisov2010anomalous,salgado2013normal}.   Despite their simplicity, these models are useful in modeling several physical systems~\cite{denisov2010biased,denisov2007analytically,denisov2007arrival,denisov2010anomalous,salgado2013normal} allowing a better  understanding of the origin of the transport properties from the very deterministic dynamics~\cite{salgado2013normal}. Moreover, these models already exhibit normal and anomalous diffusion~\cite{kunz2003mechanical,denisov2010anomalous,salgado2013normal}, and it has been proved that the emergence of such behaviors depends on the correlations of the random potentials or, in general, on the validity of the central limit theorem of a certain observable~\cite{salgado2013normal}. It is worth to point out that the deterministic models do not match with  the zero temperature limit  of those models with Gaussian disorder at finite temperature. Indeed, to our knowledge, it has not been considered the influence of Gaussian white noise in the transport properties of deterministic models (e.g., those considered in Ref.~\cite{denisov2010biased} or in Ref.~\cite{salgado2013normal}).  In this work we found that in this class of systems the presence of Gaussian white noise induces remarkably different transport properties with respect to those found at zero temperature. Examples of the latter are the non-monotonic dependence of the diffusion coefficient on the temperature over a wide range of tilt strengths and the enhancement of the diffusion coefficient by decreasing the disorder. Moreover, we are able to calculate exact expressions for the drift  and diffusion coefficients valid for arbitrary tilt strengths and noise intensities. The latter allows us to understand the origin of such properties as an interplay between the deterministic and noisy dynamics, which we explore in detail in this work.

The paper is organized as follows: In Sec.~\ref{sec:TheModel} we present the working model and establish the notation used throughout this work. In Sec.~\ref{sec:EffectiveDiffusion} we give  exact formulas for the drift and diffusion coefficients of our model. We prove that these quantities reduce to the corresponding transport coefficients for deterministic systems in the zero temperature limit. We also show that the drift and diffusion coefficients reported in this work are consistent with those given for systems without  disorder. In Sec.~\ref{sec:optimal} we test our formulas for the particle current and the diffusion coefficient for the case of uncorrelated potentials. In particular,  we study the phenomenon of enhancement of the diffusion coefficient by weakening the disorder in the polymer. We compare the results analytically obtained for such quantities and those obtained by means of Langevin dynamics simulations. Finally in Sec.~\ref{sec:Conclusions} we give a brief discussion of our results and the main conclusions of our work. Two appendices are included containing detailed calculations.

\section{Model}
\label{sec:TheModel}

We will consider an ensemble of Brownian particles with overdamped dynamics moving on a 1D disordered potential $V(x)$ subjected to an external force $F$. The equation of motion of one of these particles is given by the stochastic differential equation,
\begin{equation}
\label{eq:overdamped_particle}
\gamma dX_t =  (f(X_t) +F)dt + \varrho_0 dW_t,
\end{equation}
where $X_t$ stands for the position of the particle and $W_t$ is a standard Wiener process. The constants $\varrho_0^2$, $F$ and $\gamma $ are the noise intensity, the strength of the tilt and the friction coefficient respectively. According to  the fluctuation-dissipation theorem $\varrho_0^2 = 2 \gamma \beta^{-1}$, where $\beta$, as usual, stands for the inverse temperature times the Boltzmann constant, $\beta = 1/k_B T$. The function $f(x)$ represents minus the gradient of the potential $V(x)$ that the particle feels due to its interaction with the substrate where the motion occurs. 

\begin{figure}[t]
\begin{center}
\scalebox{0.28}{\includegraphics{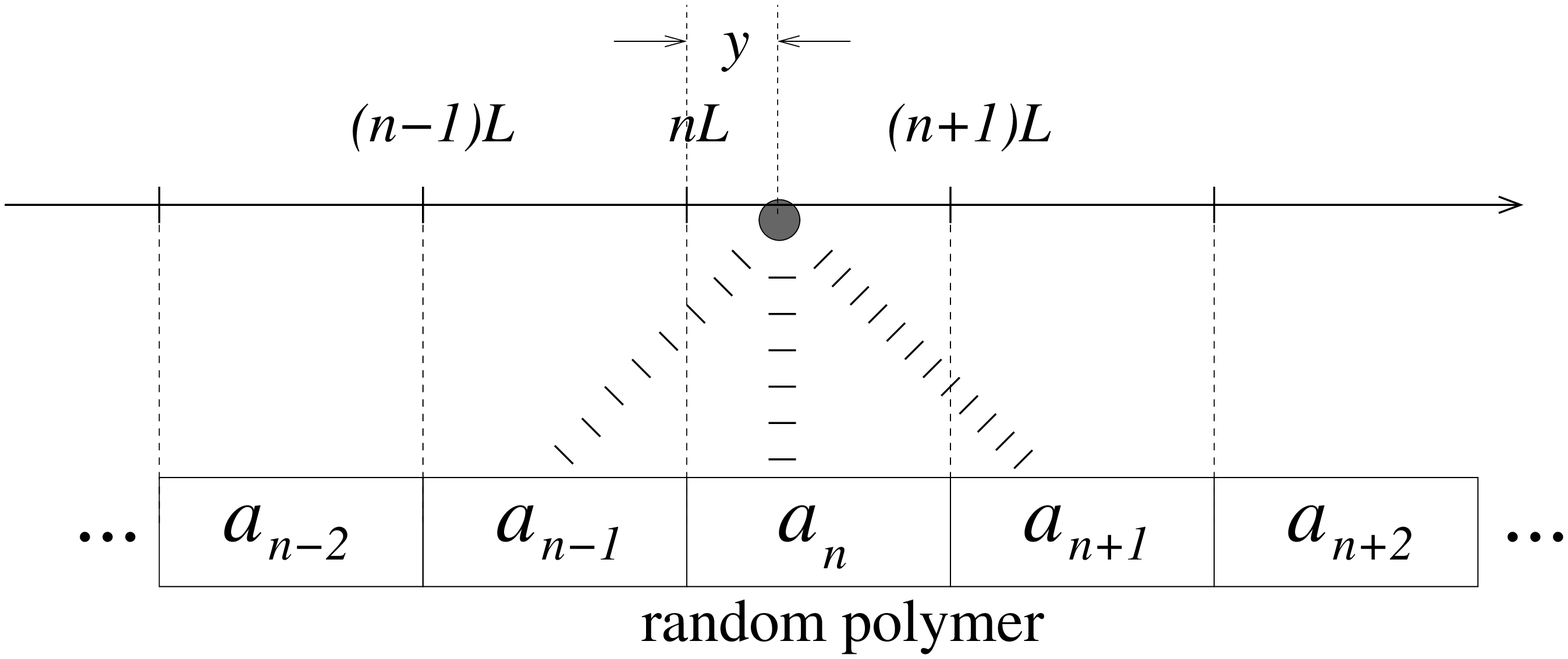}}
\end{center}
     \caption{
     Schematic representation of the particle-polymer interaction. When located at $x = nL +y$ the particle feels a potential $V(x) = \psi[y,\sigma^n(\mathbf{a})]$ that depends on the monomer type $a_n$, the relative position $y$ on the unitary cell, and on the neighbor monomer $a_{n-1}$ and $a_{n+1}$. The dependence of the interaction on the $n$th monomer and its neighbors is represented by the dependence of $\psi$ on the $n$th shift of the sequence, $\sigma^n(\mathbf{a})$. 
     }
\label{fig:particle-polymer}
\end{figure}
%

The substrate (or the polymer) on which the particles are moving will be assumed to be made up of  ``unit cells''  of constant length $L$. The unit cells represent the monomers comprising the polymer. Let us call $\mathcal{A}$ the set of possible monomer types, which can be assumed to be finite or countable infinite. Let the polymer be represented by an infinite symbolic sequence  $\mathbf{a}:=(\dots,a_{-1},a_{0},a_{1},\dots)$, where $a_j \in \mathcal{A}$ stands for the monomer type located on the $j$th cell, for all $j\in \mathbb{Z}$.  The set of possible random polymers will be denoted by $\mathcal{A}^{\mathbb{Z}}$ according to the conventional notation in symbolic dynamics~\cite{lind1995introduction}. As in Ref.~\cite{salgado2013normal}, we assume that the disordered potential $V(x)$ is the result of the interaction of a particle with the random polymer. 

Let $x\in \mathbb{R}$ be the particle position along the substrate $\mathbf{a}\in\mathcal{A}^\mathbb{Z}$. It is clear that the potential is a function of both, the position and the substrate, i.e. $V(x) = \psi(x,\mathbf{a})$. If $x=0$ we assume that the particle is located at the beginning of the $0$th monomer $a_0$. Let us write $x$  as  $x = y +nL$, where $y$ is the relative position of the particle on the $n$th cell. Then, the random potential $V(x)$ can be seen as a function $\psi$ depending on the relative position $y$, the $n$th monomer kind $a_n$ and possibly on the closest monomers to $a_n$, i.e., $a_{n-1}$ and $a_{n+1}$ (or even, depending on all the monomers in the chain if the interactions are large enough).  See Fig.~\ref{fig:particle-polymer} for an schematic representation of this situation. Let $\sigma : \mathcal{A}^\mathbb{Z} \to \mathcal{A}^\mathbb{Z} $ represents the shift mapping on $\mathcal{A}^\mathbb{Z}$, i.e., $\mathbf{b} =\sigma(\mathbf{a})$, then $b_i = a_{i+1}$ for all $i\in\mathbb{Z}$. Following the notation of Ref.~\cite{salgado2013normal}, we have that the potential at the $n$th cell can be written as
\[
V(x) = \psi[y,\sigma^{n}(\mathbf{a})].
\]
Notice that the above property for the potential $\psi$ can be generalized as follows: the displacement of the particle by an integer number of cells, say for example $nL$, is equivalent to shifting backward the polymer the same number of cells. The latter is an action achieved by the shift mapping $\sigma^{n}(\mathbf{a})$~\cite{salgado2013normal} over the symbolic sequence $\mathbf{a}$. This property can therefore be written down as
\begin{equation}
\label{eq:shift-potential}
  \psi(x+nL,\mathbf{a})  = \psi[x,\sigma^n(\mathbf{a})],
\end{equation}
for every $x\in\mathbb{R}$.

In our work, we will assume that the substrate is generated by some stochastic process.  In other words, we assume that the substrate $\mathbf{a}$ is drawn at random by some stationary measure $\mu$ on $\mathcal{A}^\mathbb{Z}$. As in Ref.~\cite{salgado2013normal} we will also assume that such a measure $\mu$ is an ergodic and shift-invariant (i.e., translationally invariant or, equivalently, $\sigma$-invariant) probability measure.

\section{The particle current and the effective diffusion coefficient}
\label{sec:EffectiveDiffusion}

Reimann \emph{et al} in Ref.~\cite{reimann2001giant} have shown that the effective diffusion coefficient can be written in a closed form if the first and second moments of the first passage time (FPT) are known exactly. An analogous   statement has been proved for deterministic overdamped particles diffusing over disordered potentials~\cite{salgado2013normal}. For the latter case, the diffusion coefficient can be written explicitly in terms of the first and second moments of the ``crossing times'' and the corresponding pair correlation function.
Here we will combine the ideas developed in Refs.~\cite{reimann2001giant} and~\cite{salgado2013normal} to give exact expressions for the particle flux and the diffusion coefficient for an ensemble of particles in disordered potentials.  

Since there are two underlying random process in our system (the Gaussian white noise and the disordered potentials), its is necessary to introduce two kinds of averages. First let us consider a realization of the polymer $\mathbf{a} \in \mathcal{A}^\mathbb{Z}$, which fixes the potential felt by a given particle. If we put an ensemble of non-interacting  Brownian  particles over such a polymer we will denote the average over this ensemble of particles as $\langle \cdot \rangle_{\mathrm{n}}$. This average will be referred to as the \emph{average with respect to the noise}. Once a certain observable has been averaged with respect to the noise, it still depends on the specific realization of the polymer $\mathbf{a}$. Thus we need to perform a second average which should be carried out over an ensemble of different realizations of the random polymer. This average is performed by using the stationary measure $\mu$ defining the process by means of which we build up the polymer. This average will be referred to as the \emph{average over the polymer ensemble} and will be denoted by $\langle \cdot \rangle_{\mathrm{p}}$. If we perform both averages we will use the notation $\langle \langle \cdot \rangle \rangle$. Additionally, we will use the notation $\mbox{Var}_\mathrm{n}(\mathcal{O})  := \langle \mathcal{O}^2 \rangle_{\mathrm{n}} - \langle \mathcal{O} \rangle_{\mathrm{n}}^2$  and $\mbox{Var}_\mathrm{p}(\mathcal{O}) := \langle \mathcal{O}^2 \rangle_{\mathrm{p}} - \langle \mathcal{O} \rangle_{\mathrm{p}}^2$ to denote the variance of the observable $\mathcal{O}$ with respect to the noise and  the polymer ensemble respectively. Along this line, $\mbox{Var}(\mathcal{O})$ will denote the variance of the observable $\mathcal{O}$ with respect to both, the noise and the polymer ensemble, i.e., $\mbox{Var}(\mathcal{O}):=\langle \langle  \mathcal{O}^2  \rangle \rangle - \langle \langle \mathcal{O} \rangle \rangle^2$.

\subsection{The particle current}

Let us consider a realization of the polymer $\mathbf{a}\in\mathcal{A}^\mathbb{Z}$. As we stated above, such a polymer induces a random potential $V(x)$ that can be written as a function of the $n$th shift of the polymer $\sigma^n(\mathbf{a})$ and the relative position $y \in [0,L]$, i.e., $V(x) = \psi[y,\sigma^n(\mathbf{a})]$, where $x=y \mod[L]$. Let $a,b \in \mathbb{R}$ be such that $a<b$ and let $\tau(a\to b)$ denote the FPT of a Brownian particle from $a$ to $b$. Then, to evaluate the particle current we need to calculate the mean FPT. It has long been known that the moments of the FPT satisfy a recurrence relation~\cite{hanggi1990reaction},
\begin{eqnarray}
\langle \tau^n(a\to b)\rangle_{\mathrm{n}} &=& n \gamma \beta \int_a^b dx \int_{-\infty}^x dy  \,\langle \tau^{n-1}(y\to b)\rangle_{\mathrm{n}}
\nonumber
\\
&\times& \exp\big( \beta \left[ V(x) - V(y)-(x-y)F  \right] \big),
\nonumber
\\
\end{eqnarray}
for $n\in\mathbb{N}$, with $\langle \tau^0(y\to b)\rangle_{\mathrm{n}} := 1$. Since the elementary cell has a fixed length $L$, we are interested in evaluating the mean first passage time from $nL$ to $(n+1)L$  (the $n$th unit cell). This quantity will be further used to evaluate the particle current. Let $T_1 := \langle \tau(nL\to (n+1)L)\rangle_{\mathrm{n}}$ be defined as the first passage time through the $n$th unit cell.  It is clear that  $T_1$ depends on the monomer closest to the particle and its neighbors, i.e., $T_1 = T_1\left[ \sigma^n(\mathbf{a})\right]$. If we calculate $T_1(\mathbf{a})$ for arbitrary $\mathbf{a} \in\mathcal{A}^\mathbb{Z}$ we can obtain $ T_1\left[ \sigma^n(\mathbf{a})\right]$ simply by shifting $n$ times the symbolic sequence $\mathbf{a}$. This makes it clear that it is enough to calculate the mean FPT through the first unit cell, $T_1(\mathbf{a})$. In Appendix~\ref{ape:FPT-1} we show that $T_1(\mathbf{a})$  can be written as,
\begin{eqnarray}
T_1(\mathbf{a}) &=& \gamma \beta \sum_{m=1}^\infty e^{-m\beta F L} q_{+}(\mathbf{a})q_{-}[\sigma^{-m}(\mathbf{a})] 
\nonumber
\\
&+& \gamma \beta \int_0^L Q_-(x,\mathbf{a}) B_+(x,\mathbf{a})dx.
\label{eq:T1}
\end{eqnarray}
Here, the functions $q_+,q_- :\mathcal{A}^\mathbb{Z} \to \mathbb{R}$ are defined as,
\begin{eqnarray}
\label{eq:def-qpm}
q_\pm (\mathbf{a} ) &=& \int_0^L dx \exp\big( \pm \beta [ \psi(x,\mathbf{a}) - xF] \big).
\end{eqnarray}
We also define the functions $B_{\pm} : \mathbb{R} \times\mathcal{A}^\mathbb{Z} \to \mathbb{R} $ and $Q_{\pm} :  \mathbb{R} \times\mathcal{A}^\mathbb{Z} \to \mathbb{R} $ as,
\begin{eqnarray}
\label{eq:def-Qpm}
Q_\pm (x,\mathbf{a}) &=& \int_0^x dy \exp\big( \pm \beta [ \psi(y,\mathbf{a}) - yF] \big),
\\
\label{eq_def-Bpm}
B_\pm (x,\mathbf{a}) &=&\exp\big( \pm \beta [ \psi(x,\mathbf{a}) - xF] \big).
\end{eqnarray}

Once we have obtained an expression for the FPT averaged with respect to the noise we need average over the polymer ensemble. This gives, 
\begin{eqnarray}
\langle T_1(\mathbf{a}) \rangle_{\mathrm{p}} &=& \gamma \beta \sum_{m=1}^\infty e^{-m\beta F L} A_q(m)  
\nonumber
\\
&+& \gamma \beta \left \langle \int_0^L Q_-(x,\mathbf{a}) B_+(x,\mathbf{a})dx \right \rangle_{\mathrm{p}},
\label{eq:<T1>}
\end{eqnarray}
where,
\[
A_q(m) := \big \langle q_{+}(\mathbf{a})q_{-}[\sigma^{-m}(\mathbf{a})] \big\rangle_{\mathrm{p}}.
\]

Now let us consider a special case to obtain a more simple expression for $\langle T_1(\mathbf{a}) \rangle_{\mathrm{p}}$.  Let the particle be located at the $n$th cell and assume that the particle-polymer  interaction is such that the random potential at $x = y + nL$ depends only on the $n$th monomer $a_n$. This is equivalent to say that, 
\begin{equation}
\label{eq:pot_one_momomer}
V(x) = \psi[x,\sigma^n(\mathbf{a})] = \psi(x,a_n).
\end{equation}
Let us assume additionally that the polymer is built up at random by means of the Bernoulli measure. This imply that the monomers in the chain are concatenated at random to comprise the polymer without any dependence on the identity of their neighbors. Thus, any random polymer obtained in this way has no correlations at two different monomer sites. To  define the Bernoulli measure it is only necessary to specify the one-monomer probabilities $\{p(a)\, : \, a \in \mathcal{A}\}$. Here $p(a)$ gives the probability to find the monomer $a\in \mathcal{A}$ along the polymer. With these hypotheses we have that,
\[
A_q(m) = \langle q_{+}(\mathbf{a})\rangle_{\mathrm{p}} \langle q_-(\mathbf{a}) \rangle_{\mathrm{p}}.
\]
Notice that the last expression no longer depends on $m$ since the Bernoulli measure is shift invariant. This allows us write the polymer average of the mean FPT as,
\begin{eqnarray}
\langle T_1 (\mathbf{a}) \rangle_{\mathrm{p}} &=& \gamma \beta \frac{ e^{-\beta F L}}{1- e^{-\beta F L}} \langle q_{+}(\mathbf{a})\rangle_{\mathrm{p}} \langle q_-(\mathbf{a}) \rangle_{\mathrm{p}}  
\nonumber
\\
&+& \gamma \beta \left \langle \int_0^L Q_-(x,\mathbf{a}) B_+(x,\mathbf{a})dx \right \rangle_{\mathrm{p}}.
\label{eq:<T1>_uncorrelated}
\end{eqnarray}

Following the arguments given in Refs.~\cite{reimann2001giant,reimann2002diffusion,salgado2013normal}, we have that the particle flux is given by,
\begin{equation}
\label{eq:Jeff-uncorrelated}
J_{\mathrm{eff}} = \frac{L}{\langle T_1 \rangle_{\mathrm{p}}}.
\end{equation}
In the following subsection we will derive with more details this expression as well as the expression for the diffusion coefficient.

\subsection{The effective diffusion coefficient}
\label{subsec:EffectiveDiffusion}

The random variable $\tau(0\to L)$ gives the time that the particle spends crossing from the left to the right throughout the first monomer $a_0$. If the potential is fixed, following the arguments in Refs.~\cite{reimann2001giant,reimann2002diffusion}, it is clear that the first passage time from $0$ to $nL$ can be written as the sum
\begin{equation}
\label{eq:FPT-sum}
\tau(0\to nL) = \sum_{m=0}^{n-1} \tau\big(mL\to (m+1)L \big).
\end{equation}
The last statement is true since we can neglect the ``backward transitions'' because they are suppressed by an exponential factor $\exp(-m\beta FL)$~\cite{reimann2001giant}. Next notice that $\tau(0\to nL)$ can be considered as a sum of random variables which are not necessarily independent. According to the central limit theorem~\cite{gnedenko1968limit,gouezel2004central,chazottes2012fluctuations} we have that a sum of random variables (appropriately normalized) converge  to a normal distribution as $n\to \infty$ if the correlations decay fast enough.  In this way, a sufficient condition for $\tau(0\to nL)$ to have an asymptotically normal distribution is that, the FPT $\tau(nL \to (n+1)L)$ has pair-correlations,
\begin{eqnarray}
C_\tau (\ell) &:=& \big\langle \big\langle \tau(0\to L) \tau\big(\ell L \to(\ell+1) L\big) \big\rangle\mathbf{\big\rangle} - 
\big \langle\big \langle \tau(0\to L) \big\rangle\big\rangle^2,
\nonumber
\end{eqnarray}
decaying faster than $\ell^{-1}$. 

The correlation of the FPT at two distant unit cells, say for example the $m$th and $l$th cells, arises only by the correlations between the monomers $a_m$ and $a_l$. This is because the noise generates no correlations between FPT's at distant sites even in the case of periodic potentials (i.e., fully correlated potentials)~\cite{reimann2001giant}. From these arguments it follows that the average (with respect to the noise)  $\langle \tau(0\to L) \tau\big(\ell L \to(\ell+1) L\big)\rangle_\mathrm{n} $ can be factorized as the product of the averages $T_1(\mathbf{a}) T_1\left[ \sigma^{\ell}(\mathbf{a}) \right]$. This allows us write the correlation function $C(\ell)$ as,
\begin{eqnarray}
C_\tau (\ell) &=&
 \big\langle T_1(\mathbf{a}) T_1\left[ \sigma^{\ell}(\mathbf{a}) \right] \big\rangle_\mathrm{p} - 
\big \langle T_1 (\mathbf{a}) \big\rangle^2_\mathrm{p} .
\end{eqnarray}

In Ref.~\cite{salgado2013normal} it is shown that the FPT $\tau(0\to nL)$, which is written as an ergodic sum in Eq.~\eqref{eq:FPT-sum}, has an asymptotic normal distribution, then the random variable $N_t$ defined implicitly by the equation
\[
\sum_{m=0}^{N_t-1} \tau\big(mL\to (m+1)L \big) = t,
\]
has an asymptotic normal distribution with mean
\begin{equation}
\label{eq:<Nt>}
\langle N_t \rangle = \frac{t}{\langle T_1 \rangle_{\mathrm{p}} },
\end{equation}
and variance
\begin{equation}
\label{eq:Var_Nt}
\mathrm{Var}(N_t) = \frac{ \varrho_\tau^2  t}{\langle T_1 \rangle_{\mathrm{p}}^3}.
\end{equation}
Here the constant $ \varrho_\tau^2$ is defined as,
\begin{eqnarray}
 \varrho_\tau^2 &:=& \big\langle \big\langle \tau^2 (0\to L) \big\rangle\big\rangle - 
\big \langle\big \langle \tau(0\to L) \big\rangle\big\rangle^2 + 2 \sum_{m=1}^\infty C_\tau(m).
\nonumber
\end{eqnarray}

If we identify the process $X_t$ (which is governed by Eq.~\eqref{eq:overdamped_particle}) with the process $N_t$ by means of the relation $X_t = L N_t$,  then the particle current, according to Eq.~\eqref{eq:<Nt>}, is given by
\[
J_{\mathrm{eff}} :=  \lim_{t \to \infty}\frac{ \langle\langle X_t \rangle\rangle}{t} = \frac{L}{\langle T_1 \rangle_{\mathrm{p}}},
\]
and that the diffusion coefficient, according to Eq.~\eqref{eq:Var_Nt},  is given by,
\begin{equation}
\label{eq:Deff}
D_{\mathrm{eff}} := \lim_{t\to \infty} \frac{ \mathrm{Var} (X_t)}{ 2t} = \frac{ L^2 \varrho_\tau^2 }{2 \langle T_1 \rangle_{\mathrm{p}}^3}.
\end{equation}

Let use rewrite the diffusion coefficient in a more convenient (and physically meaningful) form. First notice that
\begin{eqnarray}
\varrho_\tau^2 &=& \big\langle\big \langle \tau^2 (0\to L) \big \rangle \big\rangle   - \langle T_1 \rangle_\mathrm{p}^2 
+\langle T_1^2 \rangle_\mathrm{p} - \langle T_1^2 \rangle_\mathrm{p}
\nonumber
\\
&+& 2 \sum_{m=1}^\infty C_\tau(m)
\nonumber
\\
&=&\big\langle \langle \tau^2 (0\to L)  \rangle_\mathrm{n} -  \langle \tau (0\to L)  \rangle_\mathrm{n}^2  \big\rangle_\mathrm{p} 
+ \langle T_1^2 \rangle_\mathrm{p} - \langle T_1 \rangle_\mathrm{p}^2
\nonumber
\\
&+& 2 \sum_{m=1}^\infty C_\tau(m),
\nonumber
\\
\end{eqnarray}
or equivalently, 
\begin{eqnarray}
\label{eq:rho_three_parts}
\varrho_\tau^2
&=& 
\big\langle \mathrm{Var}_{\mathrm{n}} [\tau( 0\to L)] \big \rangle_\mathrm{p} + \mbox{Var}_{\mathrm{p}}(T_1) + 2 \sum_{m=1}^\infty C_\tau(m),
\nonumber
\\
\end{eqnarray}
where $\mathrm{Var}_{\mathrm{n}} [\tau( 0\to L)] :=  \langle \tau^2 (0\to L)  \rangle_\mathrm{n} -  \langle \tau (0\to L)  \rangle_\mathrm{n}^2 $ and $\mbox{Var}_{\mathrm{p}}(T_1) := \langle T_1^2 \rangle_\mathrm{p} - \langle T_1 \rangle_\mathrm{p}^2 $.
This expression for $\varrho_\tau^2$ states that the total variance of the FPT is the sum of three contributions: $i$) the average over the polymer ensemble of the variance of the FPT, i.e.,  $\big\langle \mathrm{Var}_{\mathrm{n}} [\tau( 0\to L)] \big \rangle_\mathrm{p}  $, $ii$) the variance with respect to the polymer ensemble of the mean FPT, i.e., $\mbox{Var}_{\mathrm{p}}(T_1)$ and $iii)$ the sum of the correlations of the mean FPT.  Equation~\eqref{eq:rho_three_parts} implies that the diffusion coefficient can be decomposed into two parts,
\begin{equation}
\label{eq_Deff_2parts}
D_{\mathrm{eff}} = D_{\mathrm{noisy}} + D_{\mathrm{det}},
\end{equation}
where  $D_{\mathrm{noisy}}$ and $D_{\mathrm{det}}$ will be referred to as the \emph{noisy and deterministic parts} of $D_{\mathrm{eff}}$ respectively. These quantities are defined as follows
\begin{eqnarray}
D_{\mathrm{noisy}} &=&
\frac{L^2\big\langle \mathrm{Var}_{\mathrm{n}} (\tau( 0\to L)) \big \rangle_\mathrm{p} }{ 2 \langle T_1 \rangle_{\mathrm{p}}^3 }.
\\
D_{\mathrm{det}} &=& \frac{ L^2 \mbox{Var}_{\mathrm{p}}(T_1) + 2 L^2 \sum_{m=1}^\infty C_\tau(m) }{2  \langle T_1 \rangle_{\mathrm{p}}^3}.
\end{eqnarray}
In Appendix~\ref{ape:FPT-1} we show that the mean passage time $T_1$ reduces to the corresponding ``crossing time'' in the zero temperature limit. This implies that $D_{\mathrm{det}} $ tends to the deterministic diffusion coefficient according to Ref.~\cite{salgado2013normal}. In the same limit (zero temperature) the contribution $D_{\mathrm{noisy}} $ goes to zero since the variance with respect to the noise of the FPT,  $\mbox{Var}_{\mathrm{n}} [\tau( 0\to L)]$, goes to zero as the temperature vanishes. This means  that $D_{\mathrm{eff}}$ tends to the deterministic diffusion coefficient in the limit of zero temperature.

On the other hand, when there the polymer is not disordered (for example, the case in which the polymer consist of one and only one monomer type) the variance of the mean FPT, $\mbox{Var}_{\mathrm{p}} (T_1)$, is zero. This is because the mean FPT, $T_1(\mathbf{a})$,  is no longer a random variable but a constant (a consequence of the fact that the polymer $\mathbf{a}$ is not random).  Thus $D_{ \mathrm{det} }= 0$ in this case.  Moreover, it is clear that $\big\langle \mathrm{Var}_{\mathrm{n}} [\tau( 0\to L)] \big \rangle_\mathrm{p} = \mathrm{Var}_{\mathrm{n}} [\tau( 0\to L)]$  and that $\big\langle T_1 \big \rangle_\mathrm{p} = T_1$ because  the first and the second moments of the FPT are no longer random variables as well. Therefore it is not necessary to average over a ``polymer ensemble''. This implies that $D_{ \mathrm{noisy} }$ reduces to the diffusion coefficient for periodic potentials given by  Reimann \emph{et al} in Ref.~\cite{reimann2001giant} in the ``zero disorder'' limit,
\[
D_{\mathrm{eff}} \to 
\frac{L^2 \mathrm{Var}_{\mathrm{n}} (\tau( 0\to L)) }{ 2 \langle \tau( 0\to L) \rangle_{\mathrm{n}}^3 }.
\]

All this shows that our formula for the diffusion coefficient, given by Eq.~\eqref{eq:Deff}, is consistent with the previous findings reported  in Refs.~\cite{salgado2013normal} and~\cite{reimann2001giant}.

\bigskip

\subsection{The diffusion coefficient for uncorrelated potentials}

It is clear that the main difficulty we face when we try to calculate the diffusion coefficient by means of the formula~\eqref{eq:Deff} is the evaluation of the corresponding averages. For our model we can give an explicit  expression for $D_{\mathrm{eff}}$ in the case of uncorrelated polymers.  Consider again the potential model generated by the interaction of the particle with the closest monomer to it. Thus, this potential model only depends on one monomer, or equivalently, on one ``coordinate'' of $\mathbf{a}$ (see Eq.~\eqref{eq:pot_one_momomer}). We assume that the polymer has a stationary measure defined by the Bernoulli measure described in Sect.~\ref{subsec:EffectiveDiffusion}. First notice that for the Bernoulli measure the correlation  function $C_\tau(\ell )$ vanish for all $\ell\in \mathbb{N}$.   In this way, for uncorrelated random polymers we have that the effective diffusion coefficient reduces to,
\begin{equation}
\label{eq:Deff-uncorrelated}
D_{\mathrm{eff}} =  \frac{ L^2 \big[ \big\langle \mathrm{Var}_{\mathrm{n}} [\tau( 0\to L)] \big \rangle_\mathrm{p} + \mbox{Var}_{\mathrm{p}}(T_1) \big] }{2 \langle T_1 \rangle_{\mathrm{p}}^3}.
\end{equation}

In Ref.~\cite{reimann2002diffusion}, Reimann \textit{et al} gave an expression for the second moment of the FPT. Particularly they gave an expression for the variance of the FPT, $\mathrm{Var}_{\mathrm{n}} [\tau( 0\to L)] $, which turns out to be general, i.e., for potentials which are not necessarily periodic.  According to Ref.~\cite{reimann2002diffusion},  the variance of the FPT is given by, 
\begin{eqnarray}
\mathrm{Var}_{\mathrm{n}} [\tau( 0\to L)]   &:=& \big\langle \tau^2(0\to L) \big\rangle_{\mathrm{n}} - \big\langle\tau(0\to L) \big\rangle_{\mathrm{n}}^2
\nonumber 
\\
&=& \int_0^L dx \int_{-\infty}^{x}du B_+(x,\mathbf{a})B_-(u,\mathbf{a}) 
\nonumber
\\
&\times&\mathcal{I}^2(u,\mathbf{a}).
\end{eqnarray}
In the last expression, the function $\mathcal{I}(u,\mathbf{a})$ is defined as,
\[
\mathcal{I}(u,\mathbf{a}) := \gamma\beta\, B_+(u,\mathbf{a}) \int_{-\infty}^u dz\,B_-(z,\mathbf{a}).
\]
In Appendix~\ref{ape:FPT-2} we show that $\mathrm{Var}_{\mathrm{n}} [\tau( 0\to L)] $ can be written, after some lengthly calculations, as,
\begin{widetext}
\begin{eqnarray}
\mathrm{Var}_{\mathrm{n}} [\tau( 0\to L)] &=& 
2(\gamma\beta)^2 
\sum_{n=1}^\infty
\sum_{m=1}^\infty 
\sum_{l=1}^\infty
e^{-(n+m+l)\beta F L} q_{+}(\mathbf{a}) q_{+}[\sigma^{-n}(\mathbf{a})] q_{-}[\sigma^{-m-n}(\mathbf{a})]q_{-}[\sigma^{-n-l}(\mathbf{a})] 
\nonumber
\\
&+&
4(\gamma\beta)^2  \sum_{n=1}^\infty \sum_{m=1}^\infty e^{-(n+m)\beta F L}
q_{+}(\mathbf{a}) q_{-}[\sigma^{-n-m}(\mathbf{a})] \int_0^L Q_-[x,\sigma^{-n}(\mathbf{a})] B_+[x,\sigma^{-n}(\mathbf{a})]dx
\nonumber
\\
&+&
2(\gamma\beta)^2  \sum_{n=1}^\infty e^{-n\beta F L}
q_{+}(\mathbf{a}) \int_0^L \left(Q_-[x,\sigma^{-n}(\mathbf{a})]\right)^2 B_+[x,\sigma^{-n}(\mathbf{a})]dx
\nonumber
\\
&+&
2(\gamma\beta)^2  \sum_{m=1}^\infty \sum_{l=1}^\infty e^{-(m+l)\beta F L}
q_{-}[\sigma^{-m}(\mathbf{a})] q_{-}[\sigma^{-l}(\mathbf{a})] 
\int_0^L Q_+(x,\mathbf{a}) B_+(x,\mathbf{a})dx
\nonumber
\\
&+&
4(\gamma\beta)^2   \sum_{m=1}^\infty e^{-m\beta F L}
q_{-}[\sigma^{-m}(\mathbf{a})]
\int_0^L B_+(x,\mathbf{a}) \int_0^x Q_-(u,\mathbf{a}) B_+(u,\mathbf{a})dudx
\nonumber
\\
&+&
2(\gamma\beta)^2 
\int_0^L B_+(x,\mathbf{a}) \int_0^x B_+(u,\mathbf{a}) \left[Q_-(u,\mathbf{a}) \right]^2 dudx.
\end{eqnarray}
Now, in order to evaluate the diffusion coefficient, we take the average of $\mathrm{Var}_{\mathrm{n}} [\tau( 0\to L)]$ over the polymer ensemble. In doing so, it turns out that all the sums can be done exactly. We then obtain
\begin{eqnarray}
\langle \mathrm{Var}_{\mathrm{n}} [\tau( 0\to L)] \rangle_\mathrm{p} &=& 
2(\gamma\beta)^2 \bigg\{
\frac{e^{-3\beta F L}}{(1-e^{-\beta F L})^3}
\langle q_+(\mathbf{a}) \rangle_{\mathrm{p}}^2\langle q_-(\mathbf{a}) \rangle_{\mathrm{p}}^2 
+ \bigg( \langle q_-^2(\mathbf{a}) \rangle_{\mathrm{p}} -\langle q_-(\mathbf{a}) \rangle_{\mathrm{p}}^2  \bigg) \frac{\langle q_+(\mathbf{a}) \rangle_{\mathrm{p}}^2e^{-3\beta F L}}{(1-e^{-2\beta F L})(1-e^{-\beta F L})}
\nonumber
\\
&+& 2 \frac{e^{-2\beta F L}}{\left(1-e^{-\beta F L}\right)^2} 
\langle q_+(\mathbf{a}) \rangle_{\mathrm{p}} \langle q_-(\mathbf{a}) \rangle_{\mathrm{p}}
 \langle I_0(\mathbf{a}) \rangle_{\mathrm{p}} 
 + \frac{e^{-\beta F L}}{1-e^{-\beta F L}} 
 \langle q_+(\mathbf{a}) \rangle_{\mathrm{p}} 
 \langle I_1(\mathbf{a}) \rangle_{\mathrm{p}} 
\nonumber
\\
&+& \bigg[ \frac{e^{-2\beta F L}}{\left(1-e^{-\beta F L}\right)^2}  
\langle q_-(\mathbf{a}) \rangle_{\mathrm{p}}^2 
+ \bigg( \langle q_-^2(\mathbf{a}) \rangle_{\mathrm{p}} -\langle q_-(\mathbf{a}) \rangle_{\mathrm{p}}^2  \bigg)
\frac{e^{-2\beta F L}}{1-e^{-2\beta F L}}  \bigg]\langle I_1(\mathbf{a}) \rangle_{\mathrm{p}} 
\nonumber
\\
&+&
2\frac{ e^{-\beta F L}}{1-e^{-\beta F L}}  \langle q_-(\mathbf{a}) \rangle_{\mathrm{p}} 
\langle I_3(\mathbf{a}) \rangle_{\mathrm{p}} 
+\langle I_4(\mathbf{a}) \rangle_{\mathrm{p}} \bigg\},
\label{eq:Var_n(T2)}
\end{eqnarray}
where we have defined
\begin{eqnarray}
I_0 (\mathbf{a}) &=&  \int_0^L Q_-(x,\mathbf{a}) B_+(x,\mathbf{a})dx,
\nonumber
\\
I_1(\mathbf{a}) &=& \int_0^L \left[Q_-(x,\mathbf{a})\right]^2 B_+(x,\mathbf{a})dx,
\nonumber
\\
I_2(\mathbf{a}) &=&\int_0^L Q_+(x,\mathbf{a}) B_+(x,\mathbf{a})dx,
\nonumber
\\
I_3(\mathbf{a})&=&\int_0^L B_+(x,\mathbf{a}) \int_0^x Q_-(u,\mathbf{a}) B_+(u,\mathbf{a})dudx,
\nonumber
\\
I_4(\mathbf{a})&=&\int_0^L B_+(x,\mathbf{a}) \int_0^x B_+(u,\mathbf{a}) \left[Q_-(u,\mathbf{a}) \right]^2 dudx.
\label{eq:integrals}
\end{eqnarray}

On the other hand, the variance of the mean FPT can be written down straightforwardly from Eqs.~\eqref{eq:T1} and~\eqref{eq:<T1>_uncorrelated}. Explicitly we obtain,
\begin{eqnarray}
\mbox{Var}_\mathrm{p}\left[T_1(\mathbf{a}) \right] &=& 
\beta^2 \bigg[ \frac{ e^{-2\beta F L }}{(1-e^{-\beta F L})^2}  
 \langle q_{-}(\mathbf{a})\rangle_{\mathrm{p}}^2
\bigg( \langle q_{+}^2(\mathbf{a})\rangle_{\mathrm{p}} -  \langle q_{+}(\mathbf{a})\rangle_{\mathrm{p}}^2 \bigg)
+
\beta^2 \frac{e^{-2\beta F L }}{1-e^{-2\beta F L}}  
 \langle q_{+}^2(\mathbf{a})\rangle_{\mathrm{p}} 
\bigg( \langle q_{-}^2(\mathbf{a})\rangle_{\mathrm{p}} -  \langle q_{-}(\mathbf{a})\rangle_{\mathrm{p}}^2 \bigg)
\nonumber
\\
&+&
2 \frac{e^{-2\beta F L }}{1-e^{-2\beta F L}}
 \langle q_{+}(\mathbf{a})\rangle_{\mathrm{p}} 
\bigg( \langle q_{-}(\mathbf{a}) I_0(\mathbf{a}) \rangle_{\mathrm{p}}  -  \langle q_{-}(\mathbf{a})\rangle_{\mathrm{p}}  \langle I_{0}(\mathbf{a})\rangle_{\mathrm{p}} \bigg)  
+ \langle I_{0}^2(\mathbf{a})\rangle_{\mathrm{p}} - \langle I_{0}(\mathbf{a})\rangle_{\mathrm{p}}^2 \bigg].
\end{eqnarray}

\end{widetext}

\bigskip 
\section{Optimal diffusivity}
\label{sec:optimal}

In order to test our formula for the particle current as well as for the diffusion coefficient we introduce a simple model to calculate these quantities exactly. First we will assume that the particle-polymer interaction is such that the resulting potentials have the following characteristics $i$) they rely on only one monomer (the monomer where such a particle is located) and $ii$) they are piece-wise linear.  Let $x = nL +y$ be the particle position, with $n\in\mathbb{Z}$ and $y\in [0,1]$, then we define, 
\begin{equation}
V(x) =  \left\{ \begin{array} 
            {r@{\quad \mbox{ if } \quad}l} 
   a_n y   &  0\leq  y<L/2    \\ 
   a_n (L-y )  &  L/2 \leq y < L.       \\ 
             \end{array} \right. 
\label{eq:potential-model}
\end{equation}
\begin{figure}[t]
\begin{center}
\scalebox{0.35}{\includegraphics{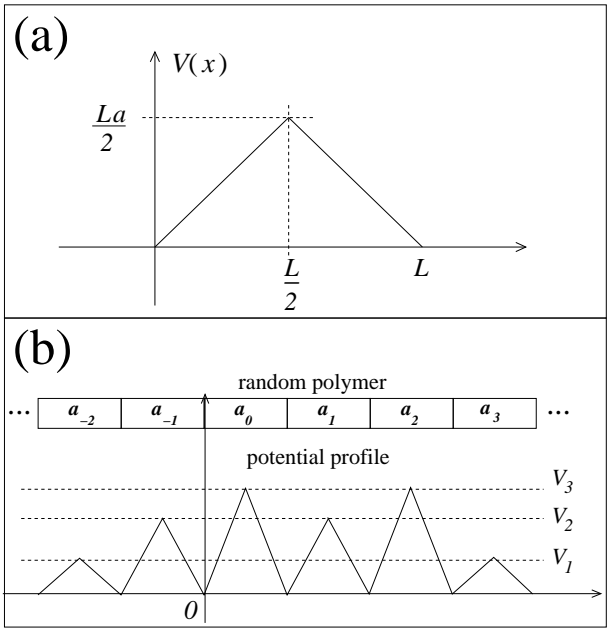}}
\end{center}
     \caption{
     Schematic representation of the potential model. (a) The potential profile on the $0$th unit cell. (b) A realization of the random potential with a few unit cells. In this case every monomer along the chain can be taken among three possible monomers ($k = 3$) with heights $V_1 = f_1 L/2$,   $V_2 = f_2 L/2$ and $V_3 = f_3 L/2$. To perform analytical calculations as well as numerical simulations we have taken the values $f_1 = 0.8$, $f_2 = 4.2$, and  $f_3 = 9.0$  (see text).
          }
\label{fig:potential-model}
\end{figure}
%
This potential model is shown schematically in Fig.~\ref{fig:potential-model}. We observe that such potential is symmetric over every unit cell, with a maximum located at $y=1/2$. The height of the potential assumed to be random taking values from a finite set. Since the height of the potential is given by $a_n L/2$ we can assume that $a_n$ represents the random variable, for every $n\in\mathbb{Z}$,  which can take values from a set $\mathcal{A} :=\{ f_j \, :\,  1\leq j \leq k \}$.  We should stress here that the sequence $\mathbf{a} := (\dots, a_{-1}, a_0, a_1,\dots) \in \mathcal{A}^\mathbb{Z} $ represents the polymer (with $a_n$ the corresponding monomers for $n\in\mathbb{Z}$). The values $f_j$  represent the ``slopes'' that can be taken by the potential and, in some way, stand for the  possible monomer types from which the polymer is built up. It is clear that the proposed potential depends only on one monomer, i.e., $V(x) = \psi(y, a_n)$ if $x = y+nL$. Since we are considering the Bernoulli measure on $\mathcal{A}^\mathbb{Z}$, we only need to specify the probability that a given monomer $a_n$  equals a monomer type $f_j$ for $1\leq j \leq k$, i.e.,  $\mathbb{P}(a_n=f_j) =: p(f_j)$.   With these quantities we can state explicitly how to average with respect to the polymer ensemble, if $h : \mathcal{A}  \to \mathbb{R}$, then we have
\begin{equation}
\label{eq:polymer_avg}
\langle h(a) \rangle_{\mathrm{p}} = \sum_{j=1}^k h(f_j) p(f_j).
\end{equation}
Notice that this average does not depends on $n$, which reflects the fact that the chosen measure is translationally invariant (or shift-invariant). 

With this potential model we have that all the integrals $q_+$, $q_-$ and $I_j$ (for $0\leq j\leq 4$) can be done exactly. This is because all the integrands appearing in these quantities have the form $e^{c x}$, with $c$ a constant. Moreover, these integrals depend only on one monomer and their polymer average can be obtained by means of the formula~\eqref{eq:polymer_avg}. The
involved integrals are obtained by using symbolic calculations in Mathematica and then numerically evaluated for the case of three monomer types.  The slopes are chosen to be $f_1 = 0.8$, $f_2 = 4.2$, and  $f_3 = 9.0$ with probabilities $p_1 := p(f_1) = 0.35$, $p_2 := p(f_2) = 0.45$, and $p_3 := p(f_3) = 0.2$. The  parameters $L$ and $\gamma $ are fixed to one.

\begin{figure}[t]
\begin{center}
\scalebox{0.33}{\includegraphics{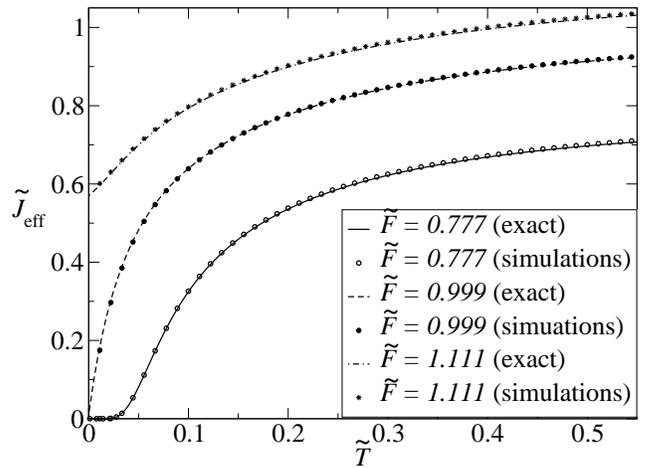}}
\end{center}
     \caption{Dimensionless particle flux as a function of the temperature. In this figure we compare the exact particle current and the corresponding obtained by means of simulations of the Langevin equation~\eqref{eq:overdamped_particle}. We display the particle current for a strength of the tilt below the critical tilt: $F= 0.777$ (solid line and open circles), near the critical tilt: $F=0.999$ (dashed line and filled circles) and above the critical tilt: $F=1.111$ (dot-dashed line and stars).
          }
\label{fig:Jeff-Theor_vs_Exp}
\end{figure}

\begin{figure}[t]
\begin{center}
\scalebox{0.33}{\includegraphics{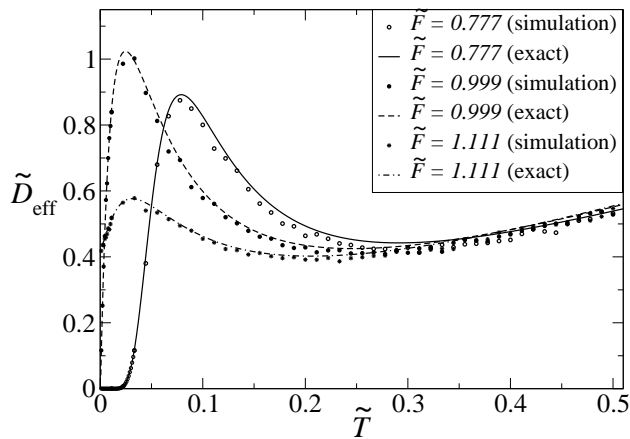}}
\end{center}
     \caption{Dimensionless diffusion coefficient as a function of the temperature. In this figure we compare the exact diffusion coefficient as a function of the temperature and the corresponding diffusion coefficient obtained by means of numerical simulations of the Langevin equation~\eqref{eq:overdamped_particle}. We display $\tilde D_\mathrm{eff}$ for $\tilde F = 0.777$ (solid line and open circles), $\tilde F = 0.999$ (dashed line filled circles) and $\tilde F = 1.111$ (dot-dashed line and stars). We appreciate that the diffusion coefficient obtained numerically fits satisfactorily (within the accuracy of our simulations) the exact diffusion coefficient for the three cases presented, below ($\tilde F = 0.777$), above ($\tilde F = 1.111$) and near critical tilt ($\tilde F = 0.999$).
               }
\label{fig:Deff-Theor_vs_Exp}
\end{figure}

In order to plot the drift and diffusion coefficients as a function of the temperature we will consider dimensionless quantities as follows: first, let $F_c$ be the \emph{critical tilt} defined as
\[
F_c := \max_{x}\{ |f(x)|\}.
\]
Next we define a dimensionless time $\tilde t = t/t_0$ with $t_0 := \gamma L/ F_c$. The dimensionless  particle current  and the dimensionless diffusion coefficient are thus defined as,
\[
\tilde J_{\mathrm{eff}} := \frac{J_{\mathrm{eff}} }{L/t_0} = \frac{\gamma J_{\mathrm{eff}}}{F_c},
\]
and
\[
\tilde D_{\mathrm{eff}} := \frac{D_{\mathrm{eff}}}{L^2/t_0} = \frac{\gamma D_{\mathrm{eff}}}{L F_c},
\]
respectively. Finally, we define the dimensionless temperature  $\tilde T$ and the dimensionless tilting force  $\tilde F$ as,
\[
\tilde T := \frac{ \beta^{-1} }{ F_c L} = \frac{k_B T}{F_c L}, 
\]
and
\[
\tilde F := \frac{ F }{ F_c }. 
\]
respectively. Notice that the dimensionless critical tilt equals one, i.e.,  $\tilde F_c = 1$.

Throughout the rest of this section we will use these dimensionless quantities ($\tilde J_{\mathrm{eff}}$, $\tilde D_{\mathrm{eff}}$, $\tilde T$ and $\tilde F$) and we will drop the ``dimensionless'' adjective to avoid unnecessary repetitions. We will make the corresponding distinctions whenever it is necessary. 

In Fig.~\ref{fig:Jeff-Theor_vs_Exp} we show the curve for the particle current obtained analytically by means of Eqs.~\eqref{eq:<T1>_uncorrelated} and~\eqref{eq:Jeff-uncorrelated} for three values of the strength of the tilt: $\tilde F=0.777$, $\tilde  F = 0.999$ and $\tilde  F =  1.111$. The same figure also displays the particle current as a function of the temperature obtained by simulating the Langevin equation~\eqref{eq:overdamped_particle} of $10 000$ particles and using an ensemble of $50$ different realizations of the random polymer. The total simulation time for every particle was $\tilde t_{\mathrm{sim}} = 18 000$. Next we obtained the corresponding average over the noise and the polymer ensemble, of the particle current $\tilde J_{\mathrm{eff}}$.  We notice a good agreement (within the accuracy of our simulations) between the theoretically predicted curves and those numerically obtained.

\begin{figure}[t]
\begin{center}
\scalebox{0.31}{\includegraphics{Fig05_a}}
\scalebox{0.31}{\includegraphics{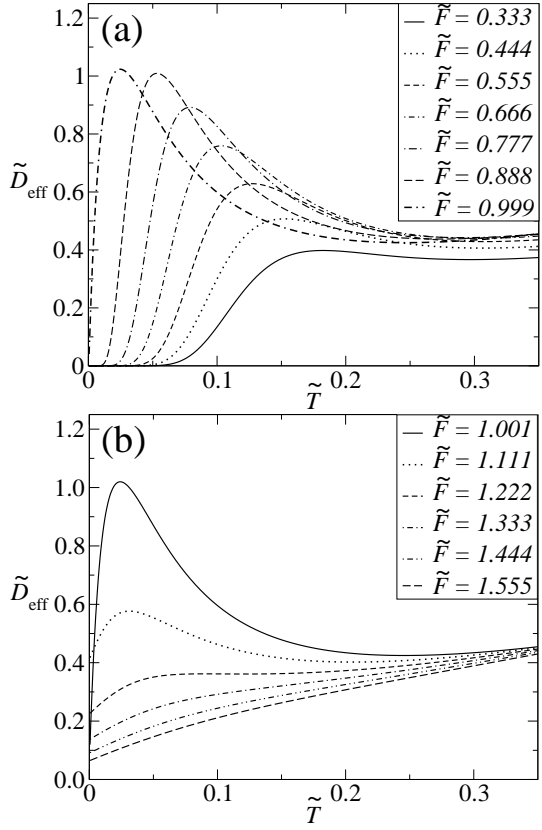}}
\end{center}
     \caption{
     Dimensionless diffusion coefficient as a function  of the temperature for strengths of the tilt (a) below the critical tilt one and (b) above the critical tilt. We observe that below and above the critical tilt the diffusion coefficient exhibits a local maximum as a function of the temperature. Below the critical tilt $\tilde F_c = 1$ we observe that the diffusion peak is more pronounced as the tilt strength increases. Once the critical tilt is reached we have the maximal diffusion peak as a function of the temperature. We also notice that, above the critical tilt, the larger tilt strength the lower diffusion peak. Indeed, above some tilt strength the diffusion peak as a function of the temperature disappears. 
          }
\label{fig:Deff-Theor}
\end{figure}

In Fig.~\ref{fig:Deff-Theor_vs_Exp} we show the curve for the diffusion coefficient predicted by our formula compared with the corresponding values obtained by means of the above described numerical simulations. We observe again a good agreement (within the accuracy of our simulations) between the theoretical and numerical curves for the three cases displayed: below ($\tilde  F=0.777$) above ($ \tilde  F=1.111$) and near ($\tilde F=0.999$) the critical tilt. It is important to stress that the diffusion coefficient has a non-trivial behavior with respect to the noise intensity. First, the diffusion coefficient increases with the noise intensity at low temperatures. Next, it reaches a local maximum at a finite temperature and then decreases as the temperature increases. Finally, the diffusivity become minimal and starts increasing again with the temperature. 
This drop in the diffusivity is, in some way, a counterintuitive phenomenon, since the dispersion of the particles is reduced while we are increasing the noise intensity. In other words, as we increase the noise strength, the particles become more ``localized'', and consequently, the transport more coherent. In Fig.~\ref{fig:Deff-Theor} we show the behavior of the diffusion coefficient as a function of the temperature (by using our exact formula) for several values of the tilting force. We can appreciate that the non-monotonous behavior of the diffusivity is a phenomenon which seems to be typical (rather than uncommon)  since it occurs for a wide range of the strength of tilt.

\begin{figure}[t]
\begin{center}
\scalebox{0.31}{\includegraphics{Fig06_a}}
\scalebox{0.31}{\includegraphics{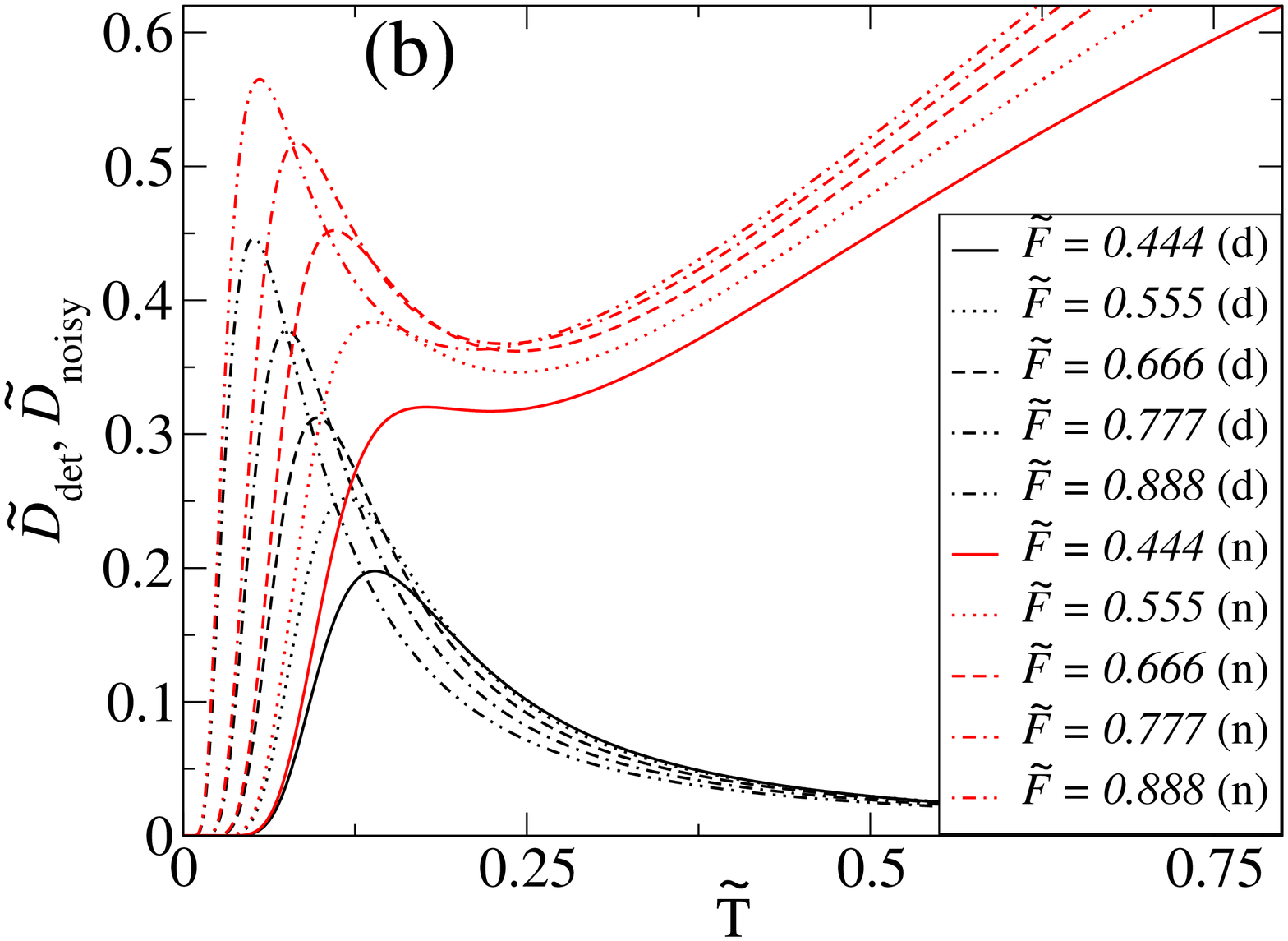}}
\end{center}
     \caption{(Color online)
Deterministic and noisy parts of the diffusion coefficient for tilting forces (a) above the critical tilt and (b) below the critical tilt. We observe that the mechanisms leading to the non-monotonous behavior of the diffusion coefficient as a function of the temperature are different in every case. Above the critical tilt, both contributions to the diffusion coefficient are monotonous. In this case the deterministic part is decreasing (black lines) while the noisy part is increasing (red lines).  These behaviors ``compete'' each other, which results in a maximum value for $\tilde D_\mathrm{eff}= \tilde D_\mathrm{det} + \tilde D_\mathrm{noisy}$ at a finite temperature. In contrast, below the critical tilt we observe that both, the deterministic and noisy contributions are already non-monotonous, a behavior which can be explained as an interplay between the escape time and a relaxation time of the system.
 }
\label{fig:Deff_noisy_vs_det}
\end{figure}
%

The non-monotonous behavior  of $\tilde  D_{\mathrm{eff}}$ is a phenomenon that has been found in a different class of systems. Previous studies reported that diffusivity exhibits this counterintuitive behavior in tilted periodic potentials~\cite{lindner2001optimal,lindner2002noise,dan2002giant,heinsalu2004diffusion}.  However, the occurrence of this phenomenon in periodic potentials is not typical at all. For example, in Ref.~\cite{lindner2001optimal,lindner2002noise} the non-monotonous behavior  was found only for potentials with a special profile. Later on, it was shown that this behavior is also found in piece-wise periodic potentials. Moreover, to observe such a phenomenon it was required  that the potential be strongly asymmetric~\cite{heinsalu2004diffusion} and it occured in a narrow window of the parameter space.  Another way to obtain the non-monotonous behavior of the diffusivity in periodic potentials is by considering an inhomogeneous friction coefficient~\cite{dan2002giant}.  In contrast, for tilted disordered potentials this behavior seems to be typical rather than unusual, as can be appreciated in Fig.~\ref{fig:Deff-Theor}. In our model, the potential profile is piece-wise constant and symmetric over every unit cell. Moreover, it has a homogeneous friction coefficient, yet the diffusion coefficient exhibits the non-monotonicity as a function of the temperature. Moreover, this behavior is more pronounced near the critical tilt and is persistent for a wide range of the tilt strengths. 

We interpret the rise and fall observed in the diffusivity as a competition between the deterministic and noisy dynamics. In Fig.~\ref{fig:Deff_noisy_vs_det} we plot the noisy and deterministic contributions to the diffusion coefficient as a function of the temperature. Above the critical tilt we observe that the deterministic part is a monotonically decreasing function of the temperature while the noisy part is an increasing one. The non-monotonicity of $\tilde  D_\mathrm{eff}$ clearly arises from the interplay between  these two behaviors. Observe that at zero temperature $\tilde D_{\mathrm{det}}$ is finite while $ \tilde D_{\mathrm{noise}}$ vanishes. As the temperature increases, $\tilde  D_{\mathrm{noise}}$ starts increasing rapidly, because there are no potential barriers in the tilted potentials. On the other hand, $\tilde D_{\mathrm{det}}$  slowly decreases becoming zero in the limit of infinite temperature. The latter occurs due to the fact that the noise ``weakens'' the interactions of the particle with the polymer. Consequently, the particles become unable to recognize the monomer type if the temperature is large enough. This implies that when the noise dominates over the deterministic dynamics the mean FPT, $T_1(\mathbf{a})$, becomes approximately the same on every unit cell. This means that the time to cross a unitary cell no longer depends on which kind of monomer the particle sees. Therefore we expect to have that the variance (with respect to the polymer ensemble) of $T_1$ be nearly zero, implying that $\tilde D_{\mathrm{det}}$ decreases as the noise increases. 

For tilts below the critical one we can appreciate that the non-monotonicity is already present for both, the deterministic and noisy parts of $\tilde D_{\mathrm{eff}}$. It is clear that in this case the diffusion coefficient is zero at zero temperature since below the critical tilt the particles cannot diffuse in absence of temperature~\cite{denisov2010biased,salgado2013normal}. Consider the case in which the strength of the tilt is slightly below the critical tilt. Thus the particle feels a potential having a set of potential wells randomly located along the polymer.  If the noise is small, the escape time is high. Due to the tilt strength, the time that the particle takes to reach a potential well once it has escaped from another is very small. Let us call such a time the ``relaxation time''. Since the relaxation time is small and the escape time large, we have an enhancement in the diffusivity at small temperatures. This is a consequence of the fact that some particles get stuck long times in the potential wells, while those that escaped from the wells rapidly move away from the particles that remain ``localized''.   If the noise intensity is further increased the escape time increases  and becomes comparable to the relaxation time. This behavior slows down the diffusivity at intermediate temperatures. If the temperature is  increased again a minimum in the diffusivity is obtained and after that it increases with the temperature. The latter is occurs because the noise fluctuations dominated over the deterministic dynamics. These behaviors clearly result in the non-monotonicity of the effective diffusion coefficient for strengths of the tilt below the critical one.

\begin{figure}[t]
\begin{center}
\scalebox{0.35}{\includegraphics{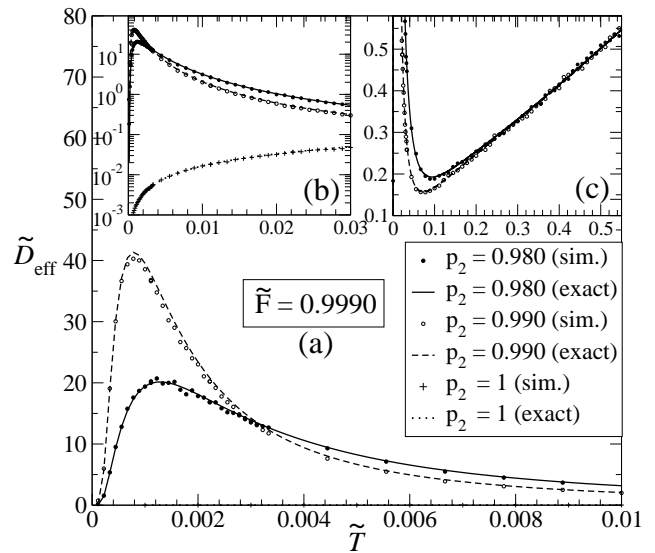}}
\end{center}
     \caption{
      Effective diffusion coefficient for weakly disordered potentials near the critical tilt. To built up the random potentials we used the set of probabilities ($i$) $p(f_1) = p(f_3) = 0.010$ and $p(f_2) = 0.980$,  ($ii$) $p(f_1) = p(f_3) = 0.005$ and $p(f_2) = 0.990$ and  ($iii$) $p(f_1) = p(f_3) = 0$ and $p(f_2) = 1$. We used the slopes $f_1 = 0.8$, $f_2 = 4.2$, $f_3 = 9.0$ 
(a) The diffusion coefficient for the set of probabilities ($i$) corresponds to the solid line (analytically calculated) and the filled circles (numerically simulated). Analogously, for the for the set of probabilities ($ii$) the diffusion coefficient corresponds to the dashed line (analytically calculated) and the open circles (numerically simulated). Within the accuracy of our numerical experiments, we observe good agreement between the simulations and the exact curves. Notice that the maximum of the diffusion coefficient at intermediate temperatures is  enhanced for these class of random potentials with weak disorder. (b) To better appreciate the diffusion coefficient for the set of of probabilities ($iii$), corresponding to a perfectly ordered polymer, we displayed the curves in a log-linear graph. (c) For large noise intensities the inset shows that after the pronounced enhancement, the diffusion coefficient increases linearly with the temperature.
        }
\label{fig:Deff-weak-disorder}
\end{figure}
%

The above explained competition between the escape time and the relaxation time becomes more pronounced at the critical tilt. Moreover, this behavior is further enhanced if we decrease the ``intensity'' of the disorder.  Consider for example a polymer consisting of one monomer kind. Assume that the potential felt by the particle is below the critical tilt and that we ``slightly''  perturb the polymer by randomly replacing some monomers.  We also assume that some of these replaced monomers induced a potential at the critical tilt. This scenario is realized if in our model we chose a set of probabilities such that $p_1$, and $p_3$ are  small and $p_2$ nearly one. With such a choice, we have that the potential is ``nearly'' periodic since the monomer $f_2$ occurs along the chain with the highest probability. The occurrence of the other two monomers is therefore considered as a ``weak disorder'' introduced in the polymer. We found that in this case the diffusion coefficient is enhanced  with respect to both, a more disordered potential and a perfectly ordered one. In Fig.~\ref{fig:Deff-weak-disorder} we plot $\tilde D_\mathrm{eff}$ as a function of the temperature for three  set of probabilities: ($i$)  $p_1 = 0.010$, $p_2 = 0.980$ and $p_3 = 0.010$, ($ii$)  $p_1 = 0.005$, $p_2 = 0.990$ and $p_3 = 0.005$ and and  ($iii$) $p(f_1) = p(f_3) = 0$ and $p(f_2) = 1$. We can appreciate how the diffusivity is enhanced as the disorder level is reduced. We also observe that the lowest diffusivity curve corresponds to the case of ``zero disorder''. Moreover, from Fig.~\ref{fig:Deff-weak-disorder} we can see that the diffusion coefficient is enhanced up to four orders of magnitude  with respect to the ``bare'' diffusivity, i.e., $ \tilde D_{\mathrm{eff}}/\tilde T \approx 6\times 10^4$.

\bigskip 
\section{Discussion and Conclusions}
\label{sec:Conclusions}

We  gave exact formulas for the particle current and the diffusion coefficient in tilted disordered potentials. We tested these formulas by means of numerical simulations of the Langevin dynamics of an ensemble of non-interacting overdamped particles sliding over  uncorrelated disordered potentials. Within the accuracy of our simulations, we found good agreement between the theoretically predicted values for these coefficients and those numerically obtained. We also found that the diffusion coefficient behaves non-monotonically with the noise intensity. Indeed, we observed that the diffusion coefficient exhibits a local maximum as a function of the temperature, a behavior which is similar to the stochastic resonance. 
We explained the occurrence of this phenomenon as a competition between the deterministic and the noisy dynamics of the system. Specifically we stated that the diffusion coefficient can be written as the sum of two contributions: $i$) the first one comes mainly from the noisy dynamics, denoted by $D_{\mathrm{noisy}}$, and which in the limit of ``zero disorder'' reduces to the usual diffusion coefficient in periodic potentials, and $ii$) a second one which comes from the deterministic dynamics of the particles on the disordered potentials, which we called $D_{\mathrm{det}}$.
We showed that the ``deterministic'' contribution reduces to the diffusion coefficient for disordered potentials for the deterministic case (given in Ref.~\cite{salgado2013normal}) in the zero temperature limit. 
Moreover, we also found that the non-monotonicity of the diffusion coefficient becomes more pronounced (and enhanced) when the disorder decreases. This enhancement is in some way analogous to the one reported by Reimann~\emph{et al}~\cite{reimann2008weak}, with the difference that the diffusion peak we reported is a function of the temperature instead a function of the strength of tilt.

\bigskip

\section*{Acknowledgements} 

The author thanks CONACyT for financial support through Grant No. CB-2012-01-183358. The author is indebted to Federico V\'azquez for carefully reading the manuscript.

\appendix
\bigskip

\section{Mean first passage time}
\label{ape:FPT-1}

In order to calculate the mean fist passage time $T_1(\mathbf{a}) := \langle \tau(0 \to L) \rangle_{\mathrm{n}}$, let us first consider the integral,
\begin{equation}
\label{eq:int_I}
\mathcal{I}(u,\mathbf{a}) := \gamma\beta\, B_+(u,\mathbf{a}) \int_{-\infty}^u dz\,B_-(z,\mathbf{a}).
\end{equation}
Remember that $B_+$ and $B_-$ are defined as,
\begin{eqnarray}
B_\pm (x,\mathbf{a}) &=& \exp\left( \pm \beta \left[V(x) - x F \right)\right].
\nonumber 
\end{eqnarray}

First let us notice that using the property~\eqref{eq:shift-potential} we obtain,
\begin{eqnarray}
B_\pm (x-nL,\mathbf{a}) &=&
\exp\left(\pm \beta \left[ \psi(x-nL,\mathbf{a} ) - (x-nL) F \right]\right)
\nonumber 
\\
&=& 
\exp\left[ \pm \beta \left(\psi[x,\sigma^{-n}(\mathbf{a})] - x F \right)\right] e^{\pm n\beta F L}
\nonumber 
\end{eqnarray}
or, equivalently,
\begin{eqnarray}
\label{eq:shift-Bpm}
B_\pm (x-nL,\mathbf{a})
&=&
B_\pm\left[x,\sigma^{-n}(\mathbf{a}) \right]e^{\pm n\beta F L}.
\end{eqnarray}
which is a property that will be used to develop further calculations.

Now, let us assume that the argument $u$ in $\mathcal{I}(u,\mathbf{a}) $ is such that $u = x-nL$ for $x\in[0,1]$.  This means that the particle is located at the $-n$th unit cell. Since $\mathcal{I}$ is defined through an integration from $-\infty$ to $u$, we can decompose it as a sum of integrals on unit cells.  This sum runs from the $-\infty$th  to the $-n$th cell, i.e., 
\begin{widetext}
\begin{eqnarray}
\mathcal{I}(x-nL,\mathbf{a}) &=&
\gamma\beta B_{+}(x-nL,\mathbf{a}) \bigg( \sum_{m=-\infty}^{-n-1} \int_{mL}^{(m+1)L} B_- (y,\mathbf{a})dy 
+\int_{-nL}^{-nL+x} B_-(y,\mathbf{a})dy
\bigg),
\nonumber
\\
&=& 
\gamma\beta B_{+}[x,\sigma^{-n}(\mathbf{a})] e^{n \beta FL} \bigg( \sum_{m=n+1}^{\infty} \int_{0}^{L} B_- (y+mL,\mathbf{a})dy 
+\int_{0}^{x} B_-(y-nL,\mathbf{a})dy
\bigg),
\nonumber
\\
&=& 
\gamma \beta B_{+}[x,\sigma^{-n}(\mathbf{a})] e^{n \beta FL} \bigg( \sum_{m=n+1}^{\infty} \int_{0}^{L}  B_-[y,\sigma^{-m}(\mathbf{a})]e^{-m\beta F L}dy 
+\int_{0}^{x} B_-[y,\sigma^{-n}(\mathbf{a})]e^{-n\beta F L}dy
\bigg),\qquad \quad
\end{eqnarray}
or, equivalently,
\begin{eqnarray}
\mathcal{I}(x-nL,\mathbf{a}) 
&=& 
\gamma\beta  \sum_{m=n+1}^{\infty} e^{(n-m) \beta FL}  B_{+}[x,\sigma^{-n}(\mathbf{a})]  q_-[\sigma^{-m}(\mathbf{a})] 
+\gamma \beta B_{+}[x,\sigma^{-n}(\mathbf{a})]  Q_-[x,\sigma^{-n}(\mathbf{a})],
\label{eq:I(x-nL)}
\end{eqnarray}
\end{widetext}
where we used the definitions of $q_\pm$, and $Q_\pm$ defined in Eqs.~\eqref{eq:def-qpm} and~\eqref{eq:def-Qpm} respectively. Since the first passage time from $x=0$ to $x=L$ is the integral of $\mathcal{I}(x-nL,\mathbf{a})$ with $n=0$, it is easy to see that,
\begin{eqnarray}
T_1(\mathbf{a}) 
&=& 
\gamma\beta  \sum_{m=1}^{\infty} e^{-m \beta FL}  q_{+}(\mathbf{a})  q_-\left[\sigma^{-m}(\mathbf{a})\right] 
\nonumber
\\
&+&
\gamma\beta \int_0^L B_{+}(x,\mathbf{a})  Q_-(x,\mathbf{a})dx,
\label{eq:T1-ape}
\end{eqnarray}
which is the result given in Eq.~\eqref{eq:T1}.

Next, we will show that $T_1(\mathbf{a})$ reduces to the ``crossing time'' given in Ref.~\cite{salgado2013normal} in the limit $\beta \to \infty$.  First notice that the tilted potential $V(x) - xF$ is always a decreasing function of $x$ if we assume that $F$ is above the critical tilt $F_c := \min_x\{-V^\prime(x)\}$. This means that the minimum of  such a tilted potential on a given interval always occurs at the upper limit and the maximum at the lower limit of the interval. These observations allow us to write down asymptotic expressions for the integrals involved in $T_1$ in the limit $\beta \to \infty$ by means of the steepest descent method. Explicitly we obtain
\begin{eqnarray}
Q_-(x,\mathbf{a}) &=& \int_0^x \exp\left( -\beta[ V(y) - yF] \right) dy,
\nonumber
\\
&\approx&
 \int_0^xe^{ -\beta[ V(x) - xF ]  -\beta [ V^\prime(x) -F](y-x) }dy,
\nonumber
\\
&=&B_-(x,\mathbf{a}) \frac{1-\exp[-\beta \phi(x,\mathbf{a})x ]}{\beta\phi(x,\mathbf{a})}.
\label{eq:Q-asymp}
\end{eqnarray}
Here we introduced the function $\phi(x,\mathbf{a}) = -V^\prime (x) + F$ as minus the gradient of the tilted potential. The function $\phi(x,\mathbf{a})$ is the total force that feels the particle at $x$ due to its interaction with the polymer $\mathbf{a}$. Notice that $\phi(x,\mathbf{a})$ is always positive if $F$ is above the critical tilt. With this result we can observe that,
\[
\int_0^LQ_-(x,\mathbf{a})  B_+(x,\mathbf{a}) dx \approx \int_0^L \frac{1-\exp[-\beta \phi(x,\mathbf{a})x ]}{\beta \phi(x,\mathbf{a})}dx,
\]
and in particular we have that,
\begin{equation}
\lim_{\beta \to \infty} \gamma \beta \int_0^LQ_-(x,\mathbf{a})  B_+(x,\mathbf{a}) dx = \gamma\int_0^L \frac{1}{\phi(x,\mathbf{a})}dx,
\label{eq:limit-tauc}
\end{equation}
Notice that the last integral coincides with the crossing time $\tau_{\mathrm{c}} : \mathcal{A}^\mathbb{Z} \to \mathbb{R}$ defined in reference~\cite{salgado2013normal},
\[ 
\tau_{\mathrm{c}}(\mathbf{a}) :=  \gamma\int_0^L \frac{1}{\phi(x,\mathbf{a})}dx.
\]

On the other hand, we have from Eq.~\eqref{eq:Q-asymp}, that
\begin{eqnarray}
q_-(\mathbf{a}) &:=&  \int_0^L \exp\left( -\beta[ V(y) - yF] \right) dy =
Q_-(L,\mathbf{a})
\nonumber
\\
&\approx&
\label{eq:q_m_asymp}
B_-(L,\mathbf{a}) \frac{1-\exp[-\beta \phi(L,\mathbf{a})L ]}{\beta \phi(L,\mathbf{a})}.
\end{eqnarray}
and similar calculations lead us to,
\begin{eqnarray}
q_+(\mathbf{a}) &:=&  \int_0^L \exp\left( \beta[ V(y) - yF] \right) dy
\nonumber
\\
&\approx&
B_+(0,\mathbf{a}) \frac{1-\exp[-\beta \phi(0,\mathbf{a})L ] }{\beta \phi(0,\mathbf{a})}.
\label{eq:q_p_asymp}
\end{eqnarray}

Now we use the properties~\eqref{eq:shift-potential} and~\eqref{eq:shift-Bpm} to obtain an asymptotic expression for  $q_-[\sigma^{-1}(\mathbf{a})]$ from Eq.~\eqref{eq:q_m_asymp}. This gives,
\begin{eqnarray}
q_-\left[\sigma^{-m}(\mathbf{a}) \right] &\approx&  B_-\left[L,\sigma^{-m}(\mathbf{a})\right] 
\frac{1-e^{-\beta \phi\left[L,\sigma^{-m}(\mathbf{a})\right]\,L }}{\beta \phi\left[L,\sigma^{-m}(\mathbf{a})\right]}
\nonumber
\\
&=&
B_-(-mL+L,\mathbf{a}) e^{ m \beta FL} 
\nonumber
\\
&\times&
\frac{1-\exp[-\beta \phi(-mL+L,\mathbf{a})L ]}{\beta \phi(-mL+L,\mathbf{a})}. \qquad
\label{eq:q_m_asymp-shifted}
\end{eqnarray}

Thus, Eq.~\eqref{eq:q_p_asymp} together with Eq.~\eqref{eq:q_m_asymp-shifted} give,
\begin{eqnarray}
q_+(\mathbf{a})  q_-\left[\sigma^{-m}(\mathbf{a}) \right] &\approx& B_+(0,\mathbf{a}) B_-(-mL+L,\mathbf{a}) 
\nonumber
\\
&\times& 
e^{ m \beta FL} \frac{1-\exp[-\beta \phi(0,\mathbf{a})L ] }{\beta \phi(0,\mathbf{a})}
\nonumber
\\
&\times&
 \frac{1-\exp[-\beta \phi(-mL+L,\mathbf{a})L ]}{\beta \phi(-mL+L,\mathbf{a})}.\qquad \quad
\end{eqnarray}

Using the fact that
\[
B_-(-mL+L,\mathbf{a}) =  \exp\left[ -\beta \psi(L-mL,\mathbf{a}) \right] e^{-(m-1)\beta F L},
\]
and that,
\[
B_+(0,\mathbf{a}) =  \exp\left[ \beta \psi(0,\mathbf{a})\right],
\]
we obtain,
\begin{eqnarray}
e^{ -m \beta FL}  q_+(\mathbf{a})  q_-\left[\sigma^{-m}(\mathbf{a}) \right] &=& 
e^{ -\beta \left[\psi(L-mL,\mathbf{a}) - \psi(0,\mathbf{a}) \right] }  
\nonumber
\\
&\times& 
  \frac{e^{-(m-1)\beta F L} }{\beta^2  \phi(0,\mathbf{a})  \phi(-mL+L,\mathbf{a}) }.
\nonumber
\end{eqnarray}
In this expression we can observe that the term $m=1$ goes to zero as $\beta^{-2}$ in the limit $\beta \to \infty$. The terms with $m>1$ decay exponentially with $\beta $ as $ e^{-(m-1)\beta F L} $. This means that the sum appearing in the expression for $T_1(\mathbf{a})$ (see Eq.~\eqref{eq:T1-ape}) vanishes in the limit of zero temperature. We have proved that the second term in $T_1(\mathbf{a})$ is finite (see Eq.~\eqref{eq:limit-tauc}) and therefore,
\[
\lim_{\beta_\to\infty} T_1(\mathbf{a}) = \tau_\mathrm{c}(\mathbf{a}).
\] 
This proves that the first passage time averaged with respect to the noise reduces to the crossing time (the ``deterministic passage time'') in the limit of zero temperature.

\bigskip
\bigskip

\section{Variance of the first passage time}
\label{ape:FPT-2}

In this Appendix we will obtain the expression~\eqref{eq:Var_n(T2)} for the variance of the FPT from the general form given by Reimann \emph{et al}~\cite{reimann2001giant,reimann2002diffusion},
\begin{eqnarray}
\mathrm{Var}_{\mathrm{n}} [\tau( 0\to L)]  &=& 2 \int_0^L dx \int_{-\infty}^{x}du\, B_+(x,\mathbf{a})B_-(u,\mathbf{a}) 
\nonumber
\\
&\times&
\mathcal{I}^2(u,\mathbf{a}). 
\end{eqnarray}
\begin{widetext}
First let us transform the integral $ \int_{-\infty}^{x}du  \dots$ as a series of integrals over unit cells as follows,
\begin{eqnarray}
\mathrm{Var}_{\mathrm{n}} [\tau( 0\to L)]  &=& 2\int_0^L dx
\sum_{n=-\infty}^{-1} \int_{nL}^{(n+1)L}du\,
B_+(x,\mathbf{a})B_-(u,\mathbf{a}) 
\mathcal{I}^2(u,\mathbf{a})
\nonumber
\\
&+&
2 \int_0^L dx \int_{0}^{x}du\,
B_+(x,\mathbf{a})B_-(u,\mathbf{a}) \mathcal{I}^2(u,\mathbf{a}).
\nonumber
\end{eqnarray}
or, equivalently
\begin{eqnarray}
\mathrm{Var}_{\mathrm{n}} [\tau( 0\to L)]  &=& 2 
\sum_{n=1}^{\infty} \int_0^L dx \int_{0}^{L}du\, B_+(x,\mathbf{a})
B_-(u-nL,\mathbf{a}) 
\mathcal{I}^2(u-nL,\mathbf{a})
+
2 \int_0^L dx \int_{0}^{x}du
B_+(x,\mathbf{a})B_-(u,\mathbf{a}) \mathcal{I}^2(u,\mathbf{a}).
\nonumber
\end{eqnarray}

Now we substitute the expression for $\mathcal{I}(u,\mathbf{a})$ given by Eq.~\eqref{eq:I(x-nL)} obtained in Appendix~\ref{ape:FPT-1}. We obtain,
\begin{eqnarray}
\mathrm{Var}_{\mathrm{n}} [\tau( 0\to L)]  &=& 2
\sum_{n=1}^{\infty} \int_0^L dx \int_{0}^{L}du\, B_+(x,\mathbf{a}) B_-(u-nL,\mathbf{a}) 
\nonumber 
\\
&\times&
\bigg[
\gamma\beta  \sum_{m=n+1}^{\infty} e^{(n-m) \beta FL}  B_{+}[u,\sigma^{-n}(\mathbf{a})]  q_-[\sigma^{-m}(\mathbf{a})] 
+\gamma \beta B_{+}[u,\sigma^{-n}(\mathbf{a})]  Q_-[u,\sigma^{-n}(\mathbf{a})]
\bigg]^2
\nonumber
\\
&+&
2 \int_0^L dx \int_{0}^{x}du
B_+(x,\mathbf{a})B_-(u,\mathbf{a}) 
\bigg[
\gamma\beta  \sum_{m=1}^{\infty} e^{-m \beta FL}  B_{+}(u,\mathbf{a})  q_-[\sigma^{-m}(\mathbf{a})] 
+\gamma \beta B_{+}(u,\mathbf{a})  Q_-(u,\mathbf{a})
\bigg]^2.
\nonumber
\end{eqnarray}
Expanding the squared terms, the above expression results in
\begin{eqnarray}
\mathrm{Var}_{\mathrm{n}} [\tau( 0\to L)]  &=& 2 \gamma^2\beta^2
\sum_{n=1}^{\infty} \int_0^L dx \int_{0}^{L}du\, B_+(x,\mathbf{a}) B_-[u,\sigma^{-n}(\mathbf{a})] e^{-n\beta FL} 
\nonumber 
\\
&\times&
\bigg[
\sum_{m=n+1}^{\infty}\sum_{l=n+1}^{\infty}
e^{(2n-m-l) \beta FL} 
B_{+}[u,\sigma^{-n}(\mathbf{a})]  q_-[\sigma^{-m}(\mathbf{a})] 
B_{+}[u,\sigma^{-n}(\mathbf{a})]  q_-[\sigma^{-l}(\mathbf{a})]
\nonumber
\\
&+&2
\sum_{m=n+1}^{\infty}
e^{(n-m) \beta FL} 
B_{+}[u,\sigma^{-n}(\mathbf{a})]  q_-[\sigma^{-m}(\mathbf{a})] 
B_{+}[u,\sigma^{-n}(\mathbf{a})]  Q_-[u,\sigma^{-n}(\mathbf{a})] 
+ 
B_{+}^2[u,\sigma^{-n}(\mathbf{a})]  Q_-^2[u,\sigma^{-n}(\mathbf{a})]
\bigg]
\nonumber
\\
&+&
 2 \gamma^2\beta^2 \int_0^L dx \int_{0}^{x}du
B_+(x,\mathbf{a})B_-(u,\mathbf{a}) 
\bigg[
\sum_{m=1}^{\infty}\sum_{l=1}^{\infty}
e^{-(m+l) \beta FL} 
B_{+}(u,\mathbf{a})  q_-[\sigma^{-m}(\mathbf{a})] 
B_{+}(u,\mathbf{a})  q_-[\sigma^{-l}(\mathbf{a})]
\nonumber
\\
&+&2
\sum_{m=1}^{\infty}
e^{-m \beta FL} 
B_{+}(u,\mathbf{a})  q_-[\sigma^{-m}(\mathbf{a})] 
B_{+}(u,\mathbf{a})  Q_-(u,\mathbf{a}) 
+ 
B_{+}^2(u,\mathbf{a})  Q_-^2(u,\mathbf{a})
\bigg],
\nonumber
\end{eqnarray}
and rearranging terms we have that
\begin{eqnarray}
\mathrm{Var}_{\mathrm{n}} [\tau( 0\to L)]  &=& 2 \gamma^2\beta^2
\sum_{n=1}^{\infty}\sum_{m=n+1}^{\infty}\sum_{l=n+1}^{\infty} 
e^{(n-m-l) \beta FL} 
\int_0^L dx \int_{0}^{L}du\, 
B_+(x,\mathbf{a}) 
 q_-[\sigma^{-m}(\mathbf{a})] 
B_{+}[u,\sigma^{-n}(\mathbf{a})]  q_-[\sigma^{-l}(\mathbf{a})]
\nonumber 
\\
&+&
4 \gamma^2\beta^2
\sum_{n=1}^{\infty} \sum_{m=n+1}^{\infty}
e^{-m \beta FL} 
\int_0^L dx \int_{0}^{L}du\, 
B_+(x,\mathbf{a}) 
q_-[\sigma^{-m}(\mathbf{a})] 
B_{+}[u,\sigma^{-n}(\mathbf{a})]  Q_-[u,\sigma^{-n}(\mathbf{a})] 
\nonumber 
\\
&+&
2 \gamma^2\beta^2
\sum_{n=1}^{\infty} e^{-n\beta F L}
\int_0^L dx \int_{0}^{L}du\, 
B_+(x,\mathbf{a}) 
B_{+}[u,\sigma^{-n}(\mathbf{a})]  Q_-^2[u,\sigma^{-n}(\mathbf{a})]
\nonumber
\\
&+&
 2 \gamma^2\beta^2 
\sum_{m=1}^{\infty}\sum_{l=1}^{\infty}
e^{-(m+l) \beta FL} 
\int_0^L dx \int_{0}^{x}du\,
B_+(x,\mathbf{a}) q_-[\sigma^{-m}(\mathbf{a})] 
B_{+}(u,\mathbf{a})  q_-[\sigma^{-l}(\mathbf{a})]
\nonumber
\\
&+&
4 \gamma^2\beta^2
\sum_{m=1}^{\infty}
e^{-m \beta FL} 
 \int_0^L dx \int_{0}^{x}du\,
B_+(x,\mathbf{a})  
q_-[\sigma^{-m}(\mathbf{a})] 
B_{+}(u,\mathbf{a})  Q_-(u,\mathbf{a}) 
\nonumber
\\
&+& 
2 \gamma^2\beta^2
 \int_0^L dx \int_{0}^{x}du\,
B_+(x,\mathbf{a})
B_{+}(u,\mathbf{a})  Q_-^2(u,\mathbf{a}).
\nonumber
\end{eqnarray}
Now, if we make the lower bound of summation over the indices $m$ and $l$ equal one  we obtain,
\begin{eqnarray}
\mathrm{Var}_{\mathrm{n}} [\tau( 0\to L)]  &=& 2 \gamma^2\beta^2
\sum_{n=1}^{\infty}\sum_{m=1}^{\infty}\sum_{l=1}^{\infty} 
e^{(-n-m-l) \beta FL}
q_-[\sigma^{-m-n}(\mathbf{a})]  q_-[\sigma^{-l-n}(\mathbf{a})]
\int_0^L dx\,
B_+(x,\mathbf{a}) 
\int_{0}^{L}du\, 
B_{+}[u,\sigma^{-n}(\mathbf{a})]  
\nonumber 
\\
&+&
4 \gamma^2\beta^2
\sum_{n=1}^{\infty} \sum_{m=1}^{\infty}
e^{-(m+n) \beta FL}
q_-[\sigma^{-m-n}(\mathbf{a})]  
\int_0^L dx\, 
B_+(x,\mathbf{a}) 
 \int_{0}^{L}du\,
 B_{+}[u,\sigma^{-n}(\mathbf{a})]  Q_-[u,\sigma^{-n}(\mathbf{a})] 
\nonumber 
\\
&+&
2 \gamma^2\beta^2
\sum_{n=1}^{\infty} e^{-n\beta F L}
\int_0^L dx\, 
B_+(x,\mathbf{a})
\int_{0}^{L}du\, 
B_{+}[u,\sigma^{-n}(\mathbf{a})]  Q_-^2[u,\sigma^{-n}(\mathbf{a})]
\nonumber
\\
&+&
 2 \gamma^2\beta^2 
\sum_{m=1}^{\infty}\sum_{l=1}^{\infty}
e^{-(m+l) \beta FL}
 q_-[\sigma^{-m}(\mathbf{a})] q_-[\sigma^{-l}(\mathbf{a})]
\int_0^L dx \int_{0}^{x}du\,
B_+(x,\mathbf{a}) 
B_{+}(u,\mathbf{a})  
\nonumber
\\
&+&
4 \gamma^2\beta^2
\sum_{m=1}^{\infty}
e^{-m \beta FL}
q_-[\sigma^{-m}(\mathbf{a})]
 \int_0^L dx \int_{0}^{x}du\,
B_+(x,\mathbf{a})   
B_{+}(u,\mathbf{a})  Q_-(u,\mathbf{a}) 
\nonumber
\\
&+& 
2 \gamma^2\beta^2
 \int_0^L dx \int_{0}^{x}du\,
B_+(x,\mathbf{a})
B_{+}(u,\mathbf{a})  Q_-^2(u,\mathbf{a}).
\nonumber
\end{eqnarray}

In the last expression we can recognize the integrals as the functions $I_j$ (for $1\leq j \leq 4$) defined in Eq.~\eqref{eq:integrals}.For uncorrelated potentials, the average over the polymer ensemble  $\langle I_j \rangle_{\mathrm{p}}$ (for $1\leq j \leq 4$) no longer depend on the summation indices. This is because the Bernoulli measure is invariant under translations along the polymer. Then we have that the summations become geometrical series which can be done exactly. After this process we arrive finally at the expression~\eqref{eq:Var_n(T2)} for the variance of the FPT.

\end{widetext}

\nocite{*}

\bibliography{EffDiffDisorderedPotentials_ref}

\begin{thebibliography}{47}%
\makeatletter
\providecommand \@ifxundefined [1]{%
 \@ifx{#1\undefined}
}%
\providecommand \@ifnum [1]{%
 \ifnum #1\expandafter \@firstoftwo
 \else \expandafter \@secondoftwo
 \fi
}%
\providecommand \@ifx [1]{%
 \ifx #1\expandafter \@firstoftwo
 \else \expandafter \@secondoftwo
 \fi
}%
\providecommand \natexlab [1]{#1}%
\providecommand \enquote  [1]{``#1''}%
\providecommand \bibnamefont  [1]{#1}%
\providecommand \bibfnamefont [1]{#1}%
\providecommand \citenamefont [1]{#1}%
\providecommand \href@noop [0]{\@secondoftwo}%
\providecommand \href [0]{\begingroup \@sanitize@url \@href}%
\providecommand \@href[1]{\@@startlink{#1}\@@href}%
\providecommand \@@href[1]{\endgroup#1\@@endlink}%
\providecommand \@sanitize@url [0]{\catcode `\\12\catcode `\$12\catcode
  `\&12\catcode `\#12\catcode `\^12\catcode `\_12\catcode `\%12\relax}%
\providecommand \@@startlink[1]{}%
\providecommand \@@endlink[0]{}%
\providecommand \url  [0]{\begingroup\@sanitize@url \@url }%
\providecommand \@url [1]{\endgroup\@href {#1}{\urlprefix }}%
\providecommand \urlprefix  [0]{URL }%
\providecommand \Eprint [0]{\href }%
\providecommand \doibase [0]{http://dx.doi.org/}%
\providecommand \selectlanguage [0]{\@gobble}%
\providecommand \bibinfo  [0]{\@secondoftwo}%
\providecommand \bibfield  [0]{\@secondoftwo}%
\providecommand \translation [1]{[#1]}%
\providecommand \BibitemOpen [0]{}%
\providecommand \bibitemStop [0]{}%
\providecommand \bibitemNoStop [0]{.\EOS\space}%
\providecommand \EOS [0]{\spacefactor3000\relax}%
\providecommand \BibitemShut  [1]{\csname bibitem#1\endcsname}%
\let\auto@bib@innerbib\@empty
\bibitem [{\citenamefont {Risken}(1984)}]{risken1984fokker}%
  \BibitemOpen
  \bibfield  {author} {\bibinfo {author} {\bibfnamefont {Hannes}\ \bibnamefont
  {Risken}},\ }\href@noop {} {\emph {\bibinfo {title} {The Fokker-Planck
  Equation}}}\ (\bibinfo  {publisher} {Springer, Berlin},\ \bibinfo {year}
  {1984})\BibitemShut {NoStop}%
\bibitem [{\citenamefont {H{\"a}nggi}\ and\ \citenamefont
  {Marchesoni}(2009)}]{hanggi2009artificial}%
  \BibitemOpen
  \bibfield  {author} {\bibinfo {author} {\bibfnamefont {P.}~\bibnamefont
  {H{\"a}nggi}}\ and\ \bibinfo {author} {\bibfnamefont {F.}~\bibnamefont
  {Marchesoni}},\ }\href@noop {} {\bibfield  {journal} {\bibinfo  {journal}
  {Rev. Mod. Phys.}\ }\textbf {\bibinfo {volume} {81}},\ \bibinfo {pages} {387}
  (\bibinfo {year} {2009})}\BibitemShut {NoStop}%
\bibitem [{\citenamefont {Brangwynne}\ \emph {et~al.}(2009)\citenamefont
  {Brangwynne}, \citenamefont {Koenderink}, \citenamefont {MacKintosh},\ and\
  \citenamefont {Weitz}}]{brangwynne2009intracellular}%
  \BibitemOpen
  \bibfield  {author} {\bibinfo {author} {\bibfnamefont {C.~P.}\ \bibnamefont
  {Brangwynne}}, \bibinfo {author} {\bibfnamefont {G.~H.}\ \bibnamefont
  {Koenderink}}, \bibinfo {author} {\bibfnamefont {F.~C.}\ \bibnamefont
  {MacKintosh}}, \ and\ \bibinfo {author} {\bibfnamefont {D.~A.}\ \bibnamefont
  {Weitz}},\ }\href@noop {} {\bibfield  {journal} {\bibinfo  {journal} {Trends
  in cell biology}\ }\textbf {\bibinfo {volume} {19}},\ \bibinfo {pages}
  {423--427} (\bibinfo {year} {2009})}\BibitemShut {NoStop}%
\bibitem [{\citenamefont {Bressloff}\ and\ \citenamefont
  {Newby}(2013)}]{bressloff2013stochastic}%
  \BibitemOpen
  \bibfield  {author} {\bibinfo {author} {\bibfnamefont {P.~C.}\ \bibnamefont
  {Bressloff}}\ and\ \bibinfo {author} {\bibfnamefont {J.~M.}\ \bibnamefont
  {Newby}},\ }\href@noop {} {\bibfield  {journal} {\bibinfo  {journal} {Rev.
  Mod. Phys.}\ }\textbf {\bibinfo {volume} {85}},\ \bibinfo {pages} {135}
  (\bibinfo {year} {2013})}\BibitemShut {NoStop}%
\bibitem [{\citenamefont {Reimann}\ \emph {et~al.}(2001)\citenamefont
  {Reimann}, \citenamefont {Van~den Broeck}, \citenamefont {Linke},
  \citenamefont {H{\"a}nggi}, \citenamefont {Rubi},\ and\ \citenamefont
  {P{\'e}rez-Madrid}}]{reimann2001giant}%
  \BibitemOpen
  \bibfield  {author} {\bibinfo {author} {\bibfnamefont {P.}~\bibnamefont
  {Reimann}}, \bibinfo {author} {\bibfnamefont {C.}~\bibnamefont {Van~den
  Broeck}}, \bibinfo {author} {\bibfnamefont {H.}~\bibnamefont {Linke}},
  \bibinfo {author} {\bibfnamefont {P.}~\bibnamefont {H{\"a}nggi}}, \bibinfo
  {author} {\bibfnamefont {J.~M.}\ \bibnamefont {Rubi}}, \ and\ \bibinfo
  {author} {\bibfnamefont {A.}~\bibnamefont {P{\'e}rez-Madrid}},\ }\href@noop
  {} {\bibfield  {journal} {\bibinfo  {journal} {Phys. Rev. Lett.}\ }\textbf
  {\bibinfo {volume} {87}},\ \bibinfo {pages} {010602} (\bibinfo {year}
  {2001})}\BibitemShut {NoStop}%
\bibitem [{\citenamefont {Reimann}\ \emph {et~al.}(2002)\citenamefont
  {Reimann}, \citenamefont {Van~den Broeck}, \citenamefont {Linke},
  \citenamefont {H{\"a}nggi}, \citenamefont {Rubi},\ and\ \citenamefont
  {P{\'e}rez-Madrid}}]{reimann2002diffusion}%
  \BibitemOpen
  \bibfield  {author} {\bibinfo {author} {\bibfnamefont {P.}~\bibnamefont
  {Reimann}}, \bibinfo {author} {\bibfnamefont {C.}~\bibnamefont {Van~den
  Broeck}}, \bibinfo {author} {\bibfnamefont {H.}~\bibnamefont {Linke}},
  \bibinfo {author} {\bibfnamefont {P.}~\bibnamefont {H{\"a}nggi}}, \bibinfo
  {author} {\bibfnamefont {J.~M.}\ \bibnamefont {Rubi}}, \ and\ \bibinfo
  {author} {\bibfnamefont {A.}~\bibnamefont {P{\'e}rez-Madrid}},\ }\href@noop
  {} {\bibfield  {journal} {\bibinfo  {journal} {Phys. Rev. E}\ }\textbf
  {\bibinfo {volume} {65}},\ \bibinfo {pages} {031104} (\bibinfo {year}
  {2002})}\BibitemShut {NoStop}%
\bibitem [{\citenamefont {Evstigneev}\ \emph {et~al.}(2008)\citenamefont
  {Evstigneev}, \citenamefont {Zvyagolskaya}, \citenamefont {Bleil},
  \citenamefont {Eichhorn}, \citenamefont {Bechinger},\ and\ \citenamefont
  {Reimann}}]{evstigneev2008diffusion}%
  \BibitemOpen
  \bibfield  {author} {\bibinfo {author} {\bibfnamefont {M.}~\bibnamefont
  {Evstigneev}}, \bibinfo {author} {\bibfnamefont {O.}~\bibnamefont
  {Zvyagolskaya}}, \bibinfo {author} {\bibfnamefont {S.}~\bibnamefont {Bleil}},
  \bibinfo {author} {\bibfnamefont {R.}~\bibnamefont {Eichhorn}}, \bibinfo
  {author} {\bibfnamefont {C.}~\bibnamefont {Bechinger}}, \ and\ \bibinfo
  {author} {\bibfnamefont {P.}~\bibnamefont {Reimann}},\ }\href@noop {}
  {\bibfield  {journal} {\bibinfo  {journal} {Phys. Rev. E}\ }\textbf {\bibinfo
  {volume} {77}},\ \bibinfo {pages} {041107} (\bibinfo {year}
  {2008})}\BibitemShut {NoStop}%
\bibitem [{\citenamefont {Lindner}\ \emph {et~al.}(2001)\citenamefont
  {Lindner}, \citenamefont {Kostur},\ and\ \citenamefont
  {Schimansky-Geier}}]{lindner2001optimal}%
  \BibitemOpen
  \bibfield  {author} {\bibinfo {author} {\bibfnamefont {B.}~\bibnamefont
  {Lindner}}, \bibinfo {author} {\bibfnamefont {M.}~\bibnamefont {Kostur}}, \
  and\ \bibinfo {author} {\bibfnamefont {L.}~\bibnamefont {Schimansky-Geier}},\
  }\href@noop {} {\bibfield  {journal} {\bibinfo  {journal} {Fluct. and Noise
  Lett.}\ }\textbf {\bibinfo {volume} {1}},\ \bibinfo {pages} {R25--R39}
  (\bibinfo {year} {2001})}\BibitemShut {NoStop}%
\bibitem [{\citenamefont {Lindner}\ and\ \citenamefont
  {Schimansky-Geier}(2002)}]{lindner2002noise}%
  \BibitemOpen
  \bibfield  {author} {\bibinfo {author} {\bibfnamefont {B.}~\bibnamefont
  {Lindner}}\ and\ \bibinfo {author} {\bibfnamefont {L.}~\bibnamefont
  {Schimansky-Geier}},\ }\href@noop {} {\bibfield  {journal} {\bibinfo
  {journal} {Phys. Rev. Lett.}\ }\textbf {\bibinfo {volume} {89}},\ \bibinfo
  {pages} {230602} (\bibinfo {year} {2002})}\BibitemShut {NoStop}%
\bibitem [{\citenamefont {Dan}\ and\ \citenamefont
  {Jayannavar}(2002)}]{dan2002giant}%
  \BibitemOpen
  \bibfield  {author} {\bibinfo {author} {\bibfnamefont {D.}~\bibnamefont
  {Dan}}\ and\ \bibinfo {author} {\bibfnamefont {A.~M.}\ \bibnamefont
  {Jayannavar}},\ }\href@noop {} {\bibfield  {journal} {\bibinfo  {journal}
  {Phys. Rev. E}\ }\textbf {\bibinfo {volume} {66}},\ \bibinfo {pages}
  {041106\_1--041106\_5} (\bibinfo {year} {2002})}\BibitemShut {NoStop}%
\bibitem [{\citenamefont {Heinsalu}\ \emph {et~al.}(2004)\citenamefont
  {Heinsalu}, \citenamefont {Tammelo},\ and\ \citenamefont
  {{\"O}rd}}]{heinsalu2004diffusion}%
  \BibitemOpen
  \bibfield  {author} {\bibinfo {author} {\bibfnamefont {E.}~\bibnamefont
  {Heinsalu}}, \bibinfo {author} {\bibfnamefont {R.}~\bibnamefont {Tammelo}}, \
  and\ \bibinfo {author} {\bibfnamefont {T.}~\bibnamefont {{\"O}rd}},\
  }\href@noop {} {\bibfield  {journal} {\bibinfo  {journal} {Phys. Rev. E}\
  }\textbf {\bibinfo {volume} {69}},\ \bibinfo {pages} {021111} (\bibinfo
  {year} {2004})}\BibitemShut {NoStop}%
\bibitem [{\citenamefont {Viovy}(2000)}]{viovy2000electrophoresis}%
  \BibitemOpen
  \bibfield  {author} {\bibinfo {author} {\bibfnamefont {J.-L.}\ \bibnamefont
  {Viovy}},\ }\href@noop {} {\bibfield  {journal} {\bibinfo  {journal} {Rev.
  Mod. Phys.}\ }\textbf {\bibinfo {volume} {72}},\ \bibinfo {pages} {813}
  (\bibinfo {year} {2000})}\BibitemShut {NoStop}%
\bibitem [{\citenamefont {Branton}(2008)}]{branton2008potential}%
  \BibitemOpen
  \bibfield  {author} {\bibinfo {author} {\bibfnamefont {D.~\textit{et al}}\
  \bibnamefont {Branton}},\ }\href@noop {} {\bibfield  {journal} {\bibinfo
  {journal} {Nat. biotech.}\ }\textbf {\bibinfo {volume} {26}},\ \bibinfo
  {pages} {1146--1153} (\bibinfo {year} {2008})}\BibitemShut {NoStop}%
\bibitem [{\citenamefont {Ashkenasy}\ \emph {et~al.}(2005)\citenamefont
  {Ashkenasy}, \citenamefont {S\'anchez-Quesada}, \citenamefont {Bayley},\ and\
  \citenamefont {Ghadiri}}]{ashkenasy2005recognizing}%
  \BibitemOpen
  \bibfield  {author} {\bibinfo {author} {\bibfnamefont {N.}~\bibnamefont
  {Ashkenasy}}, \bibinfo {author} {\bibfnamefont {J.}~\bibnamefont
  {S\'anchez-Quesada}}, \bibinfo {author} {\bibfnamefont {H.}~\bibnamefont
  {Bayley}}, \ and\ \bibinfo {author} {\bibfnamefont {M.~R.}\ \bibnamefont
  {Ghadiri}},\ }\href@noop {} {\bibfield  {journal} {\bibinfo  {journal}
  {Angewandte Chemie}\ }\textbf {\bibinfo {volume} {117}},\ \bibinfo {pages}
  {1425--1428} (\bibinfo {year} {2005})}\BibitemShut {NoStop}%
\bibitem [{\citenamefont {Slutsky}\ \emph {et~al.}(2004)\citenamefont
  {Slutsky}, \citenamefont {Kardar},\ and\ \citenamefont
  {Mirny}}]{slutsky2004diffusion}%
  \BibitemOpen
  \bibfield  {author} {\bibinfo {author} {\bibfnamefont {M.}~\bibnamefont
  {Slutsky}}, \bibinfo {author} {\bibfnamefont {M.}~\bibnamefont {Kardar}}, \
  and\ \bibinfo {author} {\bibfnamefont {L.~A.}\ \bibnamefont {Mirny}},\
  }\href@noop {} {\bibfield  {journal} {\bibinfo  {journal} {Phys. Rev. E}\
  }\textbf {\bibinfo {volume} {69}},\ \bibinfo {pages} {061903} (\bibinfo
  {year} {2004})}\BibitemShut {NoStop}%
\bibitem [{\citenamefont {Barsky}\ \emph {et~al.}(2011)\citenamefont {Barsky},
  \citenamefont {Laurence},\ and\ \citenamefont
  {Venclovas}}]{barsky2011proteins}%
  \BibitemOpen
  \bibfield  {author} {\bibinfo {author} {\bibfnamefont {D.}~\bibnamefont
  {Barsky}}, \bibinfo {author} {\bibfnamefont {T.~A.}\ \bibnamefont
  {Laurence}}, \ and\ \bibinfo {author} {\bibfnamefont {{\v{C}}.}~\bibnamefont
  {Venclovas}},\ }in\ \href@noop {} {\emph {\bibinfo {booktitle} {Biophysics of
  DNA-Protein Interactions}}}\ (\bibinfo  {publisher} {Springer},\ \bibinfo
  {year} {2011})\ pp.\ \bibinfo {pages} {39--68}\BibitemShut {NoStop}%
\bibitem [{\citenamefont {Shimamoto}(1999)}]{shimamoto1999one}%
  \BibitemOpen
  \bibfield  {author} {\bibinfo {author} {\bibfnamefont {N.}~\bibnamefont
  {Shimamoto}},\ }\href@noop {} {\bibfield  {journal} {\bibinfo  {journal} {J.
  of Biol. Chem.}\ }\textbf {\bibinfo {volume} {274}},\ \bibinfo {pages}
  {15293--15296} (\bibinfo {year} {1999})}\BibitemShut {NoStop}%
\bibitem [{\citenamefont {Gorman}\ and\ \citenamefont
  {Greene}(2008)}]{gorman2008visualizing}%
  \BibitemOpen
  \bibfield  {author} {\bibinfo {author} {\bibfnamefont {J.}~\bibnamefont
  {Gorman}}\ and\ \bibinfo {author} {\bibfnamefont {E.~C.}\ \bibnamefont
  {Greene}},\ }\href@noop {} {\bibfield  {journal} {\bibinfo  {journal} {Nature
  structural \& molecular biology}\ }\textbf {\bibinfo {volume} {15}},\
  \bibinfo {pages} {768--774} (\bibinfo {year} {2008})}\BibitemShut {NoStop}%
\bibitem [{\citenamefont {Wang}\ \emph {et~al.}(2006)\citenamefont {Wang},
  \citenamefont {Austin},\ and\ \citenamefont {Cox}}]{wang2006single}%
  \BibitemOpen
  \bibfield  {author} {\bibinfo {author} {\bibfnamefont {Y.~M.}\ \bibnamefont
  {Wang}}, \bibinfo {author} {\bibfnamefont {R.~H.}\ \bibnamefont {Austin}}, \
  and\ \bibinfo {author} {\bibfnamefont {E.~C.}\ \bibnamefont {Cox}},\
  }\href@noop {} {\bibfield  {journal} {\bibinfo  {journal} {Phys. Rev. Lett.}\
  }\textbf {\bibinfo {volume} {97}},\ \bibinfo {pages} {048302} (\bibinfo
  {year} {2006})}\BibitemShut {NoStop}%
\bibitem [{\citenamefont {Bouchaud}\ and\ \citenamefont
  {Georges}(1990)}]{bouchaud1990anomalous}%
  \BibitemOpen
  \bibfield  {author} {\bibinfo {author} {\bibfnamefont {J.-P.}\ \bibnamefont
  {Bouchaud}}\ and\ \bibinfo {author} {\bibfnamefont {A.}~\bibnamefont
  {Georges}},\ }\href@noop {} {\bibfield  {journal} {\bibinfo  {journal} {Phys.
  Rep.}\ }\textbf {\bibinfo {volume} {195}},\ \bibinfo {pages} {127--293}
  (\bibinfo {year} {1990})}\BibitemShut {NoStop}%
\bibitem [{\citenamefont {Haus}\ and\ \citenamefont
  {Kehr}(1987)}]{haus1987diffusion}%
  \BibitemOpen
  \bibfield  {author} {\bibinfo {author} {\bibfnamefont {J.~W.}\ \bibnamefont
  {Haus}}\ and\ \bibinfo {author} {\bibfnamefont {K.~W.}\ \bibnamefont
  {Kehr}},\ }\href@noop {} {\bibfield  {journal} {\bibinfo  {journal} {Phys.
  Rep.}\ }\textbf {\bibinfo {volume} {150}},\ \bibinfo {pages} {263--406}
  (\bibinfo {year} {1987})}\BibitemShut {NoStop}%
\bibitem [{\citenamefont {Havlin}\ and\ \citenamefont
  {Ben-Avraham}(1987)}]{havlin1987diffusion}%
  \BibitemOpen
  \bibfield  {author} {\bibinfo {author} {\bibfnamefont {S.}~\bibnamefont
  {Havlin}}\ and\ \bibinfo {author} {\bibfnamefont {D.}~\bibnamefont
  {Ben-Avraham}},\ }\href@noop {} {\bibfield  {journal} {\bibinfo  {journal}
  {Advances in Physics}\ }\textbf {\bibinfo {volume} {36}},\ \bibinfo {pages}
  {695--798} (\bibinfo {year} {1987})}\BibitemShut {NoStop}%
\bibitem [{\citenamefont {Sinai}(1983)}]{sinai1983limiting}%
  \BibitemOpen
  \bibfield  {author} {\bibinfo {author} {\bibfnamefont {Y.~G.}\ \bibnamefont
  {Sinai}},\ }\href@noop {} {\bibfield  {journal} {\bibinfo  {journal} {Theory
  of Probability \& Its Applications}\ }\textbf {\bibinfo {volume} {27}},\
  \bibinfo {pages} {256--268} (\bibinfo {year} {1983})}\BibitemShut {NoStop}%
\bibitem [{\citenamefont {Solomon}(1975)}]{solomon1975random}%
  \BibitemOpen
  \bibfield  {author} {\bibinfo {author} {\bibfnamefont {F.}~\bibnamefont
  {Solomon}},\ }\href@noop {} {\bibfield  {journal} {\bibinfo  {journal} {Ann.
  of Prob.}\ }\textbf {\bibinfo {volume} {3}},\ \bibinfo {pages} {1--31}
  (\bibinfo {year} {1975})}\BibitemShut {NoStop}%
\bibitem [{\citenamefont {Oshanin}\ \emph {et~al.}(1993)\citenamefont
  {Oshanin}, \citenamefont {Mogutov},\ and\ \citenamefont
  {Moreau}}]{oshanin1993steady}%
  \BibitemOpen
  \bibfield  {author} {\bibinfo {author} {\bibfnamefont {G.}~\bibnamefont
  {Oshanin}}, \bibinfo {author} {\bibfnamefont {A.}~\bibnamefont {Mogutov}}, \
  and\ \bibinfo {author} {\bibfnamefont {M.}~\bibnamefont {Moreau}},\
  }\href@noop {} {\bibfield  {journal} {\bibinfo  {journal} {J. Stat. Phys.}\
  }\textbf {\bibinfo {volume} {73}},\ \bibinfo {pages} {379--388} (\bibinfo
  {year} {1993})}\BibitemShut {NoStop}%
\bibitem [{\citenamefont {Burlatsky}\ \emph {et~al.}(1992)\citenamefont
  {Burlatsky}, \citenamefont {Oshanin}, \citenamefont {Mogutov},\ and\
  \citenamefont {Moreau}}]{burlatsky1992non}%
  \BibitemOpen
  \bibfield  {author} {\bibinfo {author} {\bibfnamefont {S.~F.}\ \bibnamefont
  {Burlatsky}}, \bibinfo {author} {\bibfnamefont {G.~S.}\ \bibnamefont
  {Oshanin}}, \bibinfo {author} {\bibfnamefont {A.~V.}\ \bibnamefont
  {Mogutov}}, \ and\ \bibinfo {author} {\bibfnamefont {M.}~\bibnamefont
  {Moreau}},\ }\href@noop {} {\bibfield  {journal} {\bibinfo  {journal} {Phys.
  Rev. A}\ }\textbf {\bibinfo {volume} {45}},\ \bibinfo {pages} {6955--6959}
  (\bibinfo {year} {1992})}\BibitemShut {NoStop}%
\bibitem [{\citenamefont {Derrida}\ and\ \citenamefont
  {Pomeau}(1982)}]{derrida1982classical}%
  \BibitemOpen
  \bibfield  {author} {\bibinfo {author} {\bibfnamefont {B.}~\bibnamefont
  {Derrida}}\ and\ \bibinfo {author} {\bibfnamefont {Y.}~\bibnamefont
  {Pomeau}},\ }\href@noop {} {\bibfield  {journal} {\bibinfo  {journal} {Phys.
  Rev. Lett.}\ }\textbf {\bibinfo {volume} {48}},\ \bibinfo {pages} {627}
  (\bibinfo {year} {1982})}\BibitemShut {NoStop}%
\bibitem [{\citenamefont {Golosov}(1984)}]{golosov1984localization}%
  \BibitemOpen
  \bibfield  {author} {\bibinfo {author} {\bibfnamefont {A.~O.}\ \bibnamefont
  {Golosov}},\ }\href@noop {} {\bibfield  {journal} {\bibinfo  {journal}
  {Commun. Math. Phys.}\ }\textbf {\bibinfo {volume} {92}},\ \bibinfo {pages}
  {491--506} (\bibinfo {year} {1984})}\BibitemShut {NoStop}%
\bibitem [{\citenamefont {Monthus}(2006)}]{monthus2006random}%
  \BibitemOpen
  \bibfield  {author} {\bibinfo {author} {\bibfnamefont {C{\'e}cile}\
  \bibnamefont {Monthus}},\ }\href@noop {} {\bibfield  {journal} {\bibinfo
  {journal} {Lett. in Math. Phys.}\ }\textbf {\bibinfo {volume} {78}},\
  \bibinfo {pages} {207--233} (\bibinfo {year} {2006})}\BibitemShut {NoStop}%
\bibitem [{\citenamefont {Monthus}(2003)}]{monthus2003localization}%
  \BibitemOpen
  \bibfield  {author} {\bibinfo {author} {\bibfnamefont {C{\'e}cile}\
  \bibnamefont {Monthus}},\ }\href@noop {} {\bibfield  {journal} {\bibinfo
  {journal} {Phys. Rev. E}\ }\textbf {\bibinfo {volume} {67}},\ \bibinfo
  {pages} {046109} (\bibinfo {year} {2003})}\BibitemShut {NoStop}%
\bibitem [{\citenamefont {Bouchaud}\ \emph {et~al.}(1990)\citenamefont
  {Bouchaud}, \citenamefont {Comtet}, \citenamefont {Georges},\ and\
  \citenamefont {Le~Doussal}}]{bouchaud1990classical}%
  \BibitemOpen
  \bibfield  {author} {\bibinfo {author} {\bibfnamefont {J.-P.}\ \bibnamefont
  {Bouchaud}}, \bibinfo {author} {\bibfnamefont {A.}~\bibnamefont {Comtet}},
  \bibinfo {author} {\bibfnamefont {A.}~\bibnamefont {Georges}}, \ and\
  \bibinfo {author} {\bibfnamefont {P.}~\bibnamefont {Le~Doussal}},\
  }\href@noop {} {\bibfield  {journal} {\bibinfo  {journal} {Annals of
  Physics}\ }\textbf {\bibinfo {volume} {201}},\ \bibinfo {pages} {285--341}
  (\bibinfo {year} {1990})}\BibitemShut {NoStop}%
\bibitem [{\citenamefont {Romero}\ and\ \citenamefont
  {Sancho}(1998)}]{romero1998brownian}%
  \BibitemOpen
  \bibfield  {author} {\bibinfo {author} {\bibfnamefont {A.~H.}\ \bibnamefont
  {Romero}}\ and\ \bibinfo {author} {\bibfnamefont {J.~M.}\ \bibnamefont
  {Sancho}},\ }\href@noop {} {\bibfield  {journal} {\bibinfo  {journal} {Phys.
  Rev. E}\ }\textbf {\bibinfo {volume} {58}},\ \bibinfo {pages} {2833--2837}
  (\bibinfo {year} {1998})}\BibitemShut {NoStop}%
\bibitem [{\citenamefont {Lopatin}\ and\ \citenamefont
  {Vinokur}(2001)}]{lopatin2001instanton}%
  \BibitemOpen
  \bibfield  {author} {\bibinfo {author} {\bibfnamefont {A.~V.}\ \bibnamefont
  {Lopatin}}\ and\ \bibinfo {author} {\bibfnamefont {V.~M.}\ \bibnamefont
  {Vinokur}},\ }\href@noop {} {\bibfield  {journal} {\bibinfo  {journal} {Phys.
  Rev. Lett.}\ }\textbf {\bibinfo {volume} {86}},\ \bibinfo {pages} {1817}
  (\bibinfo {year} {2001})}\BibitemShut {NoStop}%
\bibitem [{\citenamefont {Reimann}\ and\ \citenamefont
  {Eichhorn}(2008)}]{reimann2008weak}%
  \BibitemOpen
  \bibfield  {author} {\bibinfo {author} {\bibfnamefont {P.}~\bibnamefont
  {Reimann}}\ and\ \bibinfo {author} {\bibfnamefont {R.}~\bibnamefont
  {Eichhorn}},\ }\href@noop {} {\bibfield  {journal} {\bibinfo  {journal}
  {Phys. Rev. Lett.}\ }\textbf {\bibinfo {volume} {101}},\ \bibinfo {pages}
  {180601} (\bibinfo {year} {2008})}\BibitemShut {NoStop}%
\bibitem [{\citenamefont {Khoury}\ \emph {et~al.}(2011)\citenamefont {Khoury},
  \citenamefont {Lacasta}, \citenamefont {Sancho},\ and\ \citenamefont
  {Lindenberg}}]{khoury2011weak}%
  \BibitemOpen
  \bibfield  {author} {\bibinfo {author} {\bibfnamefont {M.}~\bibnamefont
  {Khoury}}, \bibinfo {author} {\bibfnamefont {A.~M.}\ \bibnamefont {Lacasta}},
  \bibinfo {author} {\bibfnamefont {J.~M.}\ \bibnamefont {Sancho}}, \ and\
  \bibinfo {author} {\bibfnamefont {K.}~\bibnamefont {Lindenberg}},\
  }\href@noop {} {\bibfield  {journal} {\bibinfo  {journal} {Phys. Rev. Lett.}\
  }\textbf {\bibinfo {volume} {106}},\ \bibinfo {pages} {090602} (\bibinfo
  {year} {2011})}\BibitemShut {NoStop}%
\bibitem [{\citenamefont {Pryadko}\ and\ \citenamefont
  {Lin}(2005)}]{pryadko2005driven}%
  \BibitemOpen
  \bibfield  {author} {\bibinfo {author} {\bibfnamefont {L.~P.}\ \bibnamefont
  {Pryadko}}\ and\ \bibinfo {author} {\bibfnamefont {J.-X.}\ \bibnamefont
  {Lin}},\ }\href@noop {} {\bibfield  {journal} {\bibinfo  {journal} {Phys.
  Rev. E}\ }\textbf {\bibinfo {volume} {72}},\ \bibinfo {pages} {011108}
  (\bibinfo {year} {2005})}\BibitemShut {NoStop}%
\bibitem [{\citenamefont {Kunz}\ \emph {et~al.}(2003)\citenamefont {Kunz},
  \citenamefont {Livi},\ and\ \citenamefont
  {S{\"u}t{\H{o}}}}]{kunz2003mechanical}%
  \BibitemOpen
  \bibfield  {author} {\bibinfo {author} {\bibfnamefont {H.}~\bibnamefont
  {Kunz}}, \bibinfo {author} {\bibfnamefont {R.}~\bibnamefont {Livi}}, \ and\
  \bibinfo {author} {\bibfnamefont {A.}~\bibnamefont {S{\"u}t{\H{o}}}},\
  }\href@noop {} {\bibfield  {journal} {\bibinfo  {journal} {Phys. Rev. E}\
  }\textbf {\bibinfo {volume} {67}},\ \bibinfo {pages} {011102} (\bibinfo
  {year} {2003})}\BibitemShut {NoStop}%
\bibitem [{\citenamefont {Denisov}\ \emph {et~al.}(2010)\citenamefont
  {Denisov}, \citenamefont {Denisova},\ and\ \citenamefont
  {Kantz}}]{denisov2010biased}%
  \BibitemOpen
  \bibfield  {author} {\bibinfo {author} {\bibfnamefont {S.~I.}\ \bibnamefont
  {Denisov}}, \bibinfo {author} {\bibfnamefont {E.~S.}\ \bibnamefont
  {Denisova}}, \ and\ \bibinfo {author} {\bibfnamefont {H.}~\bibnamefont
  {Kantz}},\ }\href@noop {} {\bibfield  {journal} {\bibinfo  {journal} {The
  Eur. Phys. J. B}\ }\textbf {\bibinfo {volume} {76}},\ \bibinfo {pages}
  {1--11} (\bibinfo {year} {2010})}\BibitemShut {NoStop}%
\bibitem [{\citenamefont {Denisov}\ \emph
  {et~al.}(2007{\natexlab{a}})\citenamefont {Denisov}, \citenamefont {Kostur},
  \citenamefont {Denisova},\ and\ \citenamefont
  {H{\"a}nggi}}]{denisov2007analytically}%
  \BibitemOpen
  \bibfield  {author} {\bibinfo {author} {\bibfnamefont {S.~I.}\ \bibnamefont
  {Denisov}}, \bibinfo {author} {\bibfnamefont {M.}~\bibnamefont {Kostur}},
  \bibinfo {author} {\bibfnamefont {E.~S.}\ \bibnamefont {Denisova}}, \ and\
  \bibinfo {author} {\bibfnamefont {P.}~\bibnamefont {H{\"a}nggi}},\
  }\href@noop {} {\bibfield  {journal} {\bibinfo  {journal} {Phys. Rev. E}\
  }\textbf {\bibinfo {volume} {75}},\ \bibinfo {pages} {061123} (\bibinfo
  {year} {2007}{\natexlab{a}})}\BibitemShut {NoStop}%
\bibitem [{\citenamefont {Denisov}\ \emph
  {et~al.}(2007{\natexlab{b}})\citenamefont {Denisov}, \citenamefont {Kostur},
  \citenamefont {Denisova},\ and\ \citenamefont
  {H{\"a}nggi}}]{denisov2007arrival}%
  \BibitemOpen
  \bibfield  {author} {\bibinfo {author} {\bibfnamefont {S.~I.}\ \bibnamefont
  {Denisov}}, \bibinfo {author} {\bibfnamefont {M.}~\bibnamefont {Kostur}},
  \bibinfo {author} {\bibfnamefont {E.~S.}\ \bibnamefont {Denisova}}, \ and\
  \bibinfo {author} {\bibfnamefont {P.}~\bibnamefont {H{\"a}nggi}},\
  }\href@noop {} {\bibfield  {journal} {\bibinfo  {journal} {Phys. Rev. E}\
  }\textbf {\bibinfo {volume} {76}},\ \bibinfo {pages} {031101} (\bibinfo
  {year} {2007}{\natexlab{b}})}\BibitemShut {NoStop}%
\bibitem [{\citenamefont {Denisov}\ and\ \citenamefont
  {Kantz}(2010)}]{denisov2010anomalous}%
  \BibitemOpen
  \bibfield  {author} {\bibinfo {author} {\bibfnamefont {S.~I.}\ \bibnamefont
  {Denisov}}\ and\ \bibinfo {author} {\bibfnamefont {H.}~\bibnamefont
  {Kantz}},\ }\href@noop {} {\bibfield  {journal} {\bibinfo  {journal} {Phys.
  Rev. E}\ }\textbf {\bibinfo {volume} {81}},\ \bibinfo {pages} {021117}
  (\bibinfo {year} {2010})}\BibitemShut {NoStop}%
\bibitem [{\citenamefont {Salgado-Garc{\'\i}a}\ and\ \citenamefont
  {Maldonado}(2013)}]{salgado2013normal}%
  \BibitemOpen
  \bibfield  {author} {\bibinfo {author} {\bibfnamefont {R.}~\bibnamefont
  {Salgado-Garc{\'\i}a}}\ and\ \bibinfo {author} {\bibfnamefont
  {C.}~\bibnamefont {Maldonado}},\ }\href@noop {} {\bibfield  {journal}
  {\bibinfo  {journal} {Phys. Rev. E}\ }\textbf {\bibinfo {volume} {88}},\
  \bibinfo {pages} {062143} (\bibinfo {year} {2013})}\BibitemShut {NoStop}%
\bibitem [{\citenamefont {Lind}(1995)}]{lind1995introduction}%
  \BibitemOpen
  \bibfield  {author} {\bibinfo {author} {\bibfnamefont {Douglas~Alan}\
  \bibnamefont {Lind}},\ }\href@noop {} {\emph {\bibinfo {title} {An
  Introduction to Symbolic Dynamics and Coding}}}\ (\bibinfo  {publisher}
  {Cambridge University Press, Cambridge},\ \bibinfo {year} {1995})\BibitemShut
  {NoStop}%
\bibitem [{\citenamefont {H{\"a}nggi}\ \emph {et~al.}(1990)\citenamefont
  {H{\"a}nggi}, \citenamefont {Talkner},\ and\ \citenamefont
  {Borkovec}}]{hanggi1990reaction}%
  \BibitemOpen
  \bibfield  {author} {\bibinfo {author} {\bibfnamefont {P.}~\bibnamefont
  {H{\"a}nggi}}, \bibinfo {author} {\bibfnamefont {P.}~\bibnamefont {Talkner}},
  \ and\ \bibinfo {author} {\bibfnamefont {M.}~\bibnamefont {Borkovec}},\
  }\href@noop {} {\bibfield  {journal} {\bibinfo  {journal} {Rev. Mod. Phys.}\
  }\textbf {\bibinfo {volume} {62}},\ \bibinfo {pages} {251} (\bibinfo {year}
  {1990})}\BibitemShut {NoStop}%
\bibitem [{\citenamefont {Gnedenko}\ and\ \citenamefont
  {Kolmogorov}(1968)}]{gnedenko1968limit}%
  \BibitemOpen
  \bibfield  {author} {\bibinfo {author} {\bibfnamefont {B.~V.}\ \bibnamefont
  {Gnedenko}}\ and\ \bibinfo {author} {\bibfnamefont {A.~N.}\ \bibnamefont
  {Kolmogorov}},\ }\href@noop {} {\emph {\bibinfo {title} {Limit Distributions
  for Sums of Independent Random Variables}}},\ Vol.\ \bibinfo {volume} {233}\
  (\bibinfo  {publisher} {Addison-Wesley Reading},\ \bibinfo {year}
  {1968})\BibitemShut {NoStop}%
\bibitem [{\citenamefont {Gou{\"e}zel}(2004)}]{gouezel2004central}%
  \BibitemOpen
  \bibfield  {author} {\bibinfo {author} {\bibfnamefont {S{\'e}bastien}\
  \bibnamefont {Gou{\"e}zel}},\ }\href@noop {} {\bibfield  {journal} {\bibinfo
  {journal} {Prob. Theor. and Rel. Fields}\ }\textbf {\bibinfo {volume}
  {128}},\ \bibinfo {pages} {82--122} (\bibinfo {year} {2004})}\BibitemShut
  {NoStop}%
\bibitem [{\citenamefont {Chazottes}(2012)}]{chazottes2012fluctuations}%
  \BibitemOpen
  \bibfield  {author} {\bibinfo {author} {\bibfnamefont {J.-R.}\ \bibnamefont
  {Chazottes}},\ }\bibfield  {title} {\enquote {\bibinfo {title} {Fluctuations
  of observables in dynamical systems: from limit theorems to concentration
  inequalities},}\ }\href@noop {} {\bibfield  {journal} {\bibinfo  {journal}
  {arXiv preprint. arXiv:1201.3833}\ } (\bibinfo {year} {2012})}\BibitemShut
  {NoStop}%
\end{thebibliography}%

\end{document}